\documentclass[journal]{IEEEtran}
\ifCLASSINFOpdf
  \usepackage[pdftex]{graphicx}
  \DeclareGraphicsExtensions{.pdf,.jpeg,.png}
\else
\fi
\usepackage{amsmath,bm}
\usepackage{fixmath}
\usepackage{amssymb}
\usepackage{textcomp}

\usepackage{subfig}
\usepackage{gensymb}
\usepackage{textcomp}

\begin{document}

%
\title{Intra-Pulse Polyphase Coding System for Second Trip Suppression in a Weather Radar}
%
%
%

\author{Mohit~Kumar,~\IEEEmembership{Student Member,~IEEE}, V~Chandrasekar,~\IEEEmembership{Fellow,~IEEE}}

%
%

\markboth{}%
{Shell \MakeLowercase{\textit{et al.}}: Bare Demo of IEEEtran.cls for IEEE Journals}
%



\maketitle

\begin{abstract}
This paper describes the design and implementation of intra-pulse polyphase codes for a weather radar system. Algorithms to generate codes with good correlation properties are discussed. Thereafter, a new design framework is described, which optimizes the polyphase code and corresponding mismatched filter, using a cost/error function, especially for weather radars.  It establishes the performance of these intra-pulse techniques with specific application towards second trip removal. The developed code is implemented on NASA D3R, which is a dual-frequency, dual-polarization, Doppler weather radar system. 
\end{abstract}


%
\IEEEpeerreviewmaketitle

\section{Introduction} \label{Intro}
%
%
%
%
\IEEEPARstart{T}{he} use of phase codes to radars and communication has gained a lot of momentum recently because of the new computational approaches and more computing power available to search for long phase codes. There are more degrees of freedom for the design of a polyphase code than a binary code or chirp waveform. But the design complexity is generally higher, and many computational techniques can be used to obtain the desired properties for radar application (\cite{Deng2004}, \cite{Song2016} and \cite{Haohe2009}).\par

Many phase coding schemes exist in literature to retrieve weather radar products from the second, third and so on, trip echoes (after the unambiguous range). These codes can be broadly categorized into inter-pulse codes, which may be defined as a single phase offset per pulse (to the basic waveform). In this category, random phase codes and systematic phase codes, are popular to retrieve first and overlaid second trip (\cite{Bharadwaj2007}, \cite{Zrnic1999} and many references within). The idea is to incorporate a random phase or a phase from Chu code sequence \cite{Chu1972}, per pulse, so that the second trip is whitened or gets modulated differently from the first trip or vice versa (depending upon the trip that is being retrieved). The suppression effect can be seen over the integration duration, comprising of many such pulses. Under the same category are Walsh-Hadamard codes, that have been shown in \cite{Chandrasekar2009}, to give a wide separation between co-polar and cross-polar returns (in simultaneous transmission of both polarizations). On the other hand, intra-pulse scheme rely on polyphase or binary codes within the pulse, to achieve the separation on account of its correlation properties. This orthogonality between intra-pulse polyphase codes has been used in this paper for obtaining second trip suppression, if alternate pulses are coded with these codes from the orthogonal code set.\par

Additionally, the suppression abilities of second trip echoes in case of inter-pulse phase coding schemes (discussed in \cite{Bharadwaj2007} and \cite{Zrnic1999}), are dependent upon spectral width. This is because of filtering out second trip power amongst multiple replicas. Since there is no concept of using notch filters to suppress second trips in case of intra-pulse polyphase code, because this ability is inherent in the code, thus performance is not limited by multi-modal distributions or wider spectral width in general.\par 

The properties that we would try to optimize for intra-pulse polyphase codes are the aperiodic auto-correlation for a given code, and the aperiodic cross-correlation between any pair of codes in the code set. Let us call this as an orthogonal set and begin with two polyphase codes $\{s_{1}(t),s_{2}(t)\}$, each of length $L$. For each time delay $\tau$ (time variable being $t$), this set should satisfy the following properties:
\begin{equation} \label{eq_2}
\int_{t} s_{1}(t)s_{2}^{*}(t+\tau)dt = 0, \tau = 0,1,2,...,L-1,
\end{equation}

\begin{equation}
\int_{t} s_{l}(t)s_{l}^{*}(t+\tau)dt=\begin{cases}
1, & \text{if $\tau = 0, l\in {1,2}$}.\\
0, & \text{otherwise}.
\end{cases}
\end{equation}

It conveys that the correlation between two codes in the set is zero for all lags, and the auto-correlation is an impulse like function with unity response at zero lag and vanishes at all other lags. Such polyphase codes are very difficult to design and only optimal approximations can be achieved with pseudo-orthogonality. The optimization is carried out minimizing the cost function (also known as error function) derived from the above properties over the set of possible phase values, usually with gradient search. The orthogonality between codes manifests itself in their cross-correlation function and has been utilized in our paper for suppression of second trip echoes, if alternate pulses have the same code. If more code-filter pairs could be synthesized, then suppression of other trip echoes can also be achieved.\par
The cost function (to be minimized) is the total energy in the sidelobes, better known as the integrated sidelobe level (ISL). In radar terminology, the aperiodic auto-correlation function is referred to as matched filtering. However, it is shown in \cite{Bharadwaj2012}, that the peak sidelobe level can be very effectively reduced using a mismatched filter which has a slight loss in signal to noise ratio (SNR), termed as mismatched filter loss. The correlation of code $\textbf{a}$ with mismatched filter $\textbf{b}$, can be written as:

\begin{equation}
c_{m} = \varSigma_{i=1}^{L} a_{i}b^{*}_{i+m} , m\in(-L,L), 
\end{equation}
where $L$ is length of mismatched filter (the code is zero padded to make it equal to the length $L$). Total sidelobe energy is defined as:

\begin{equation}
E = \varSigma_{m=-L+1}^{L-1} \lvert c_{m} \rvert^{2}, m \neq 0.
\end{equation}

Because this energy function is non-convex mainly due to the unimodular constraint (constant modulus waveform), the local solvers would be driven towards local minima. Thus, the problem must be approached by driving multiple local solvers throughout the search space. We start by forming the energy (i.e., error) function for a mismatched-filter-based code to optimize auto-correlation sidelobe energy. Since the minimum ISL mismatched filter is known for a given polyphase code \cite{Griep1995}, we use that closed-form solution and iterate over different codes from the search space using local and global solvers, to obtain a code-filter pair with minimum ISL. A mismatched filter can optimally reduce the peak auto-correlation sidelobe, the same idea could be extended to include cross-correlation sidelobe energy as well. Therefore, we extend this strategy to the orthogonal code set, forming the error function comprising of auto-correlation sidelobe and cross-correlation sidelobe energy. For a given pair of polyphase code, the closed form of minimum ISL filter (mismatched) is derived and then we iterate over possible code filter pairs to find optimal solution. Thus we are trying to jointly optimize code and filters in this framework to find best polyphase codes satisfying the properties in equations 1 and 2. \par

Much of the literature published in the past deals with the design of polyphase code using a matched filter approach. This either requires a high pulse width or high bandwidth, to obtain a large number of samples essential for reasonable peak auto-correlation and cross-correlation sidelobes. An increase in pulse width would lead to more blind range. A high bandwidth requirement complicates the overall system design (RF, IF, filter design etc). Because of this, we formulated a mismatched-filter-based polyphase code pulse compression system, where the pulse width and bandwidth, can be maintained relatively small compared to the filter length. Further, lesser is the ratio of sampled pulse width to the filter length, there is likely improvement in peak sidelobe level. In D3R, pulse width is decided based on the total energy required to meet the sensitivity goal and lesser blind range. This type of joint formulation utilizing polyphase codes using mismatched filter to reduce auto-correlation and cross-correlation peak sidelobes for a weather radar is quite unique. Our approach is to estimate polyphase codes, using global optimization routines to arrive at a pair of polyphase codes with corresponding mismatched filters. And the criterion for optimal solution is to achieve a minimum of combined auto-and cross-correlation sidelobe energy.
The idea behind the use of mismatched filter is since it is able to lower peak auto-correlation sidelobe, then it should be able to lower peak cross-correlation sidelobe as well (by spreading out sidelobe energy into a larger coefficient space). This idea has not been explored yet and is the novelty of this framework.  \par
The gradient of the error function helps in the minimization process with speedy convergence, saving the solver with approximation using finite difference methods. When possible, this gradient can be set to zero to obtain a closed-form solution or it can be used iteratively in algorithms such as gradient descent, to find an optimal solution. In this paper, we have derived the gradient of the error function based on mismatched filter (similar to \cite{Baden2015}) so as to aid in convergence.\par
The performance of the synthesized code is analyzed through the use of ambiguity function, which has been modified for mismatched filter and also cross-ambiguity is introduced for orthogonal waveform set. The cross-ambiguity function quantifies the performance of cross-correlation between a pair of polyphase codes, over delay-doppler plane. The ambiguity function used here was restricted to the doppler experienced by weather scenarios. Both auto- and cross-ambiguity functions are very useful to evaluate the performance of the synthesized code. \par

The generated set of polyphase codes can be used for second trip suppression. Let’s say, we design two polyphase sequences which are pseudo-orthogonal. The first and second trip echoes are coded with this sequence set. Then, the cross-correlation function between sequences in the set will give the separation of first and second trip echoes. The effectiveness of these codes, for second trip suppression will be demonstrated using simulation of weather echoes (of first and second trips) and later with D3R radar data (currently at CSU-CHILL radar site in Greeley, CO), on which this scheme has been realized. \par

The various sections in this paper are organized as follows: section \ref{section_2} starts with an existing matched-filter-based polyphase code design technique and later the mismatched-filter-based code design framework is introduced for a weather radar. This section also describes constraints, choice of an initial value, a derivative of the error function, for setting up the global optimization problem. Next, the simulation of weather echoes to test the efficacy of new codes is elaborated in section \ref{section_3}, followed by the implementation and observations from D3R radar in section \ref{section_4}.

\section{Polyphase Code Design Problem} \label{section_2}

\subsection{Matched-Filter-Based Orthogonal Polyphase Code:}\label{section1}
\begin{figure}[!t]
	\centering
	\includegraphics[width=3.5in]{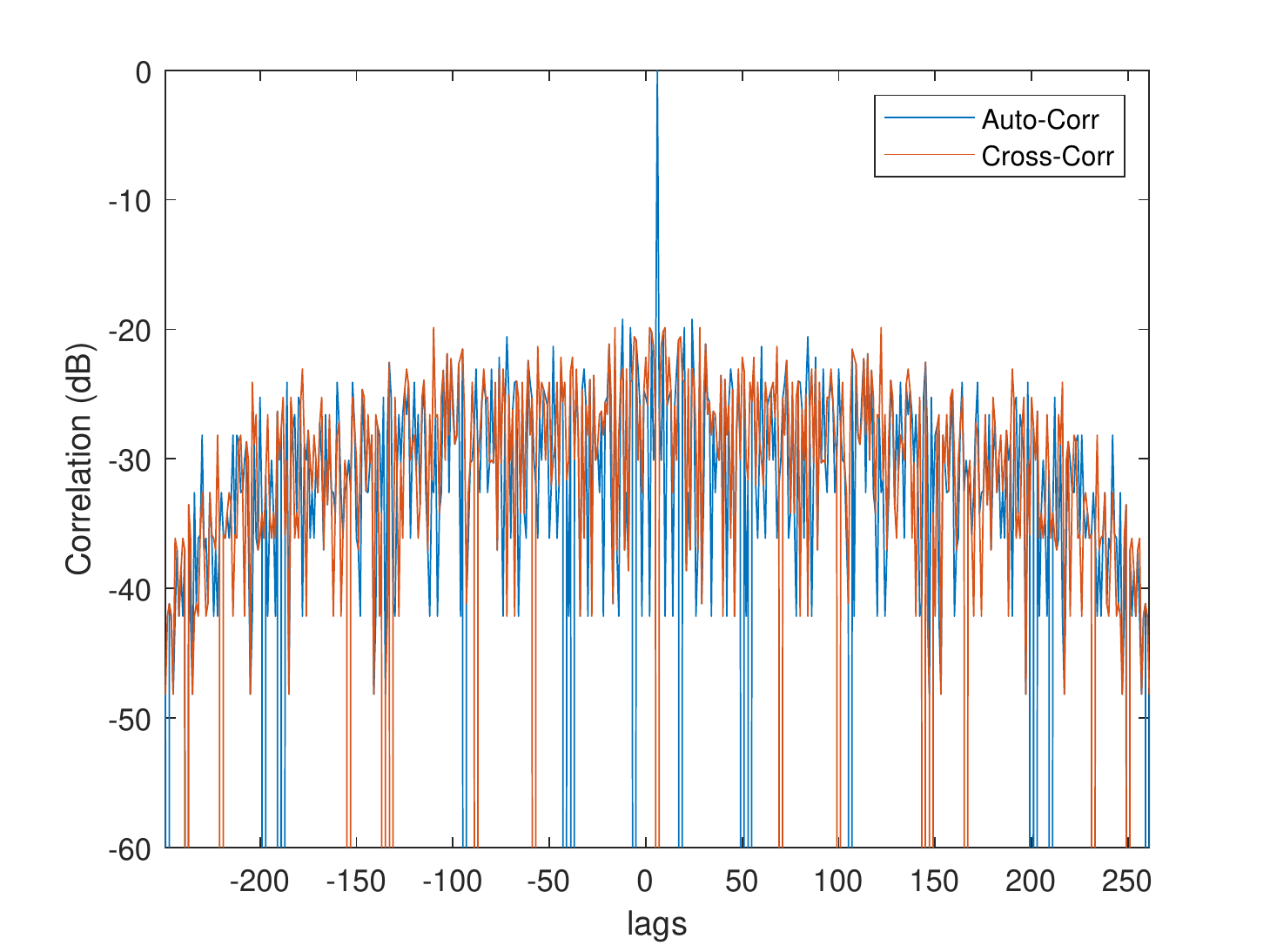}
	\caption{The correlation function (auto- and cross-correlations functions) with Hadamard codes obtained from the rows of the hadamard matrix. The code length and the filter length are both 256 samples.}
	\label{fig_sim1}
\end{figure}

\begin{figure}[!t]
	\centering
	\includegraphics[width=3.5in]{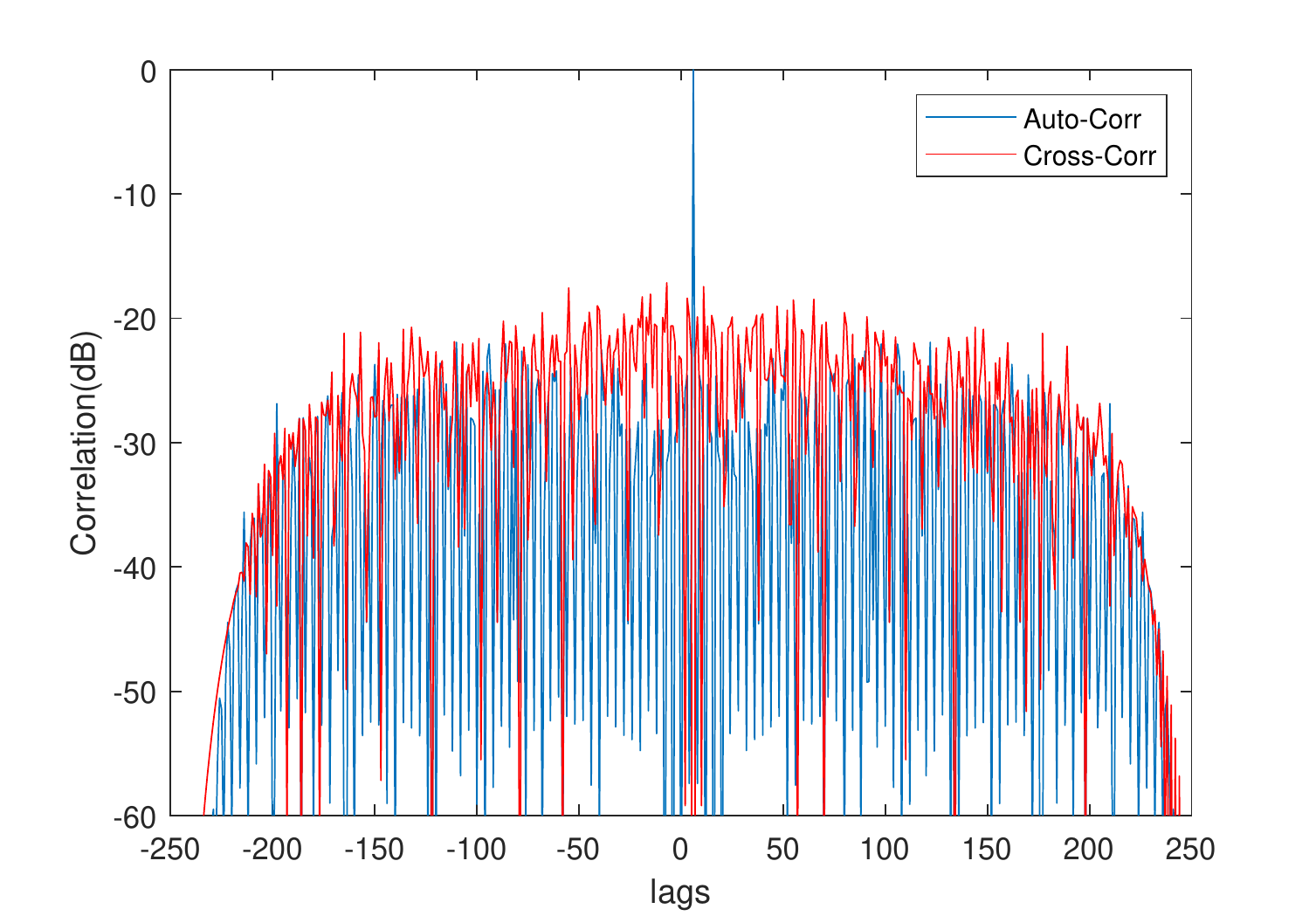}
	\caption{The auto- and cross-correlation functions with Chu codes.}
	\label{fig_sim2}
\end{figure}

\begin{figure}[!t]
	\centering
	\includegraphics[width=3.5in]{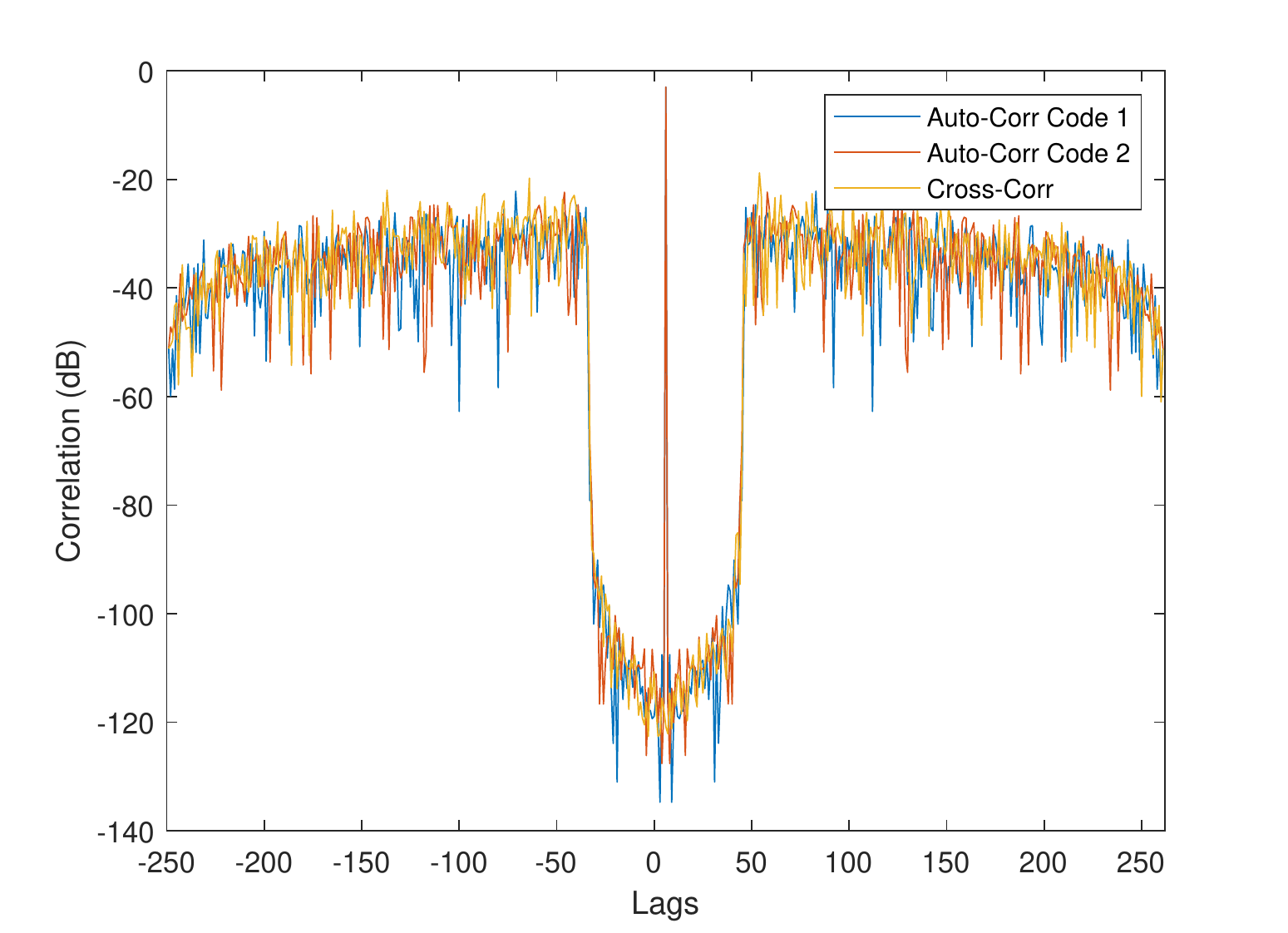}
	\caption{The correlation function with CAN based codes. Here the orthogonal sequence set comprises of \{code1, code2\}.}
	\label{fig_sim3_1}
\end{figure}

The design, in this sub-section, is based on the assumption that the filter and samples in the transmit waveform are of the same length and matched to the code transmitted. Matched filter optimizes the signal to noise ratio of the target, with the Gaussian noise assumption, while the mismatched filter optimizes the signal to sidelobe ratio, with large lengths. There are number of papers (\cite{Haohe2009}, \cite{Deng2004}) which elaborate or optimize the correlation function. But the majority of them either use large pulse widths or higher bandwidth. The comparison of different codes and their correlation function is given in Fig. \ref{fig_sim1}, \ref{fig_sim2}, \ref{fig_sim3_1}, based on Hadamard, Chu and Cyclic Algorithms (CA) respectively. These are all with a matched filter assumption. \par
The Hadamard codes are known for their orthogonal properties. These codes can be obtained by selecting a row of Hadamard Matrix $\textbf{M}_{L}$, which is a $L \times L$ matrix. Each row of this matrix differs from other rows in exactly $L/2$ places. Additionally, $\textbf{M}_{L}\textbf{M}_{L}^{t} = L\textbf{I}_{L}$, any two rows of this matrix are orthogonal at zero correlation lag, but we need to see the response at other lags as well, which is given in the Fig. \ref{fig_sim1}. For this simulation, the Hadamard code was generated using the recursive formula:

\begin{equation}
\textbf{M}_{L} = 
\begin{pmatrix} 
\textbf{M}_{L/2} & \textbf{M}_{L/2} \\
\textbf{M}_{L/2} & \overline{\textbf{M}_{L/2}}
\end{pmatrix},
\end{equation} 

where $\overline{\textbf{M}_{L/2}}$ is the complement of $\textbf{M}_{L/2}$. When $L=2$, the Hadamard code is:

\begin{equation}
\textbf{M}_{2} = 
\begin{bmatrix} 
1 & 1 \\
1 & -1
\end{bmatrix}.
\end{equation}  

It is easy to observe from Fig. \ref{fig_sim1} that the correlation function value is non-zero for all lags and the peak sidelobe is around $20dB$ below the mainlobe. 
 
Systematic Chu codes are of the type given by, $a = exp(j \bm{\psi})$ \cite{Chu1972}. These are designed in such a way that the phase change on a sub-pulse basis should be:

\begin{equation}\label{eq_4}
\bm{\phi}_{l} = \bm{\psi}_{l-1} - \bm{\psi}_{l} = n\pi l^{2}/L ;\qquad l = 0,...,L-1.
\end{equation}
The integer $L$ is made equal to the number of samples, which forces the code to repeat through one or more full cycles in the sample sequence. Depending on the choice of $n$, the periodicity of the code is $L$ or sub-multiples of $L$. For more details on the construction of these codes, the reader is referred to \cite{Zrnic1999} and \cite{Chu1972}. This Chu code has zero auto-correlation for all lags except 0 and multiples of $L$, and can be obtained by selecting $n$ relatively prime to $L$. For demonstration, we select $n = 1$ and $L = 256$, so that there is one auto-correlation peak in the convolution integral of equation \eqref{eq_2}. The two Chu codes are obtained by selecting an overall length of 512 and then using one half of the sequence as one code (and the other half as the second code). If we plot the cross correlation function of these two codes (shown in Fig. \ref{fig_sim2}), we can see that there is a deep null in the cross-correlation function at the place of auto-correlation peak. But to achieve this level of performance, the code length has to be large. \par

The polyphase orthogonal signal design can also be carried out using cyclic algorithms (CA) and CA-New (CAN) (introduced in \cite{Haohe2009}). In that paper, it is shown that to design good polyphase code waveform sets with good auto- and cross-correlation properties, we can minimize the following error function, shown to be equivalent to minimizing the sidelobe energy:

\begin{equation}
\varepsilon = ||\textbf{R}_{0} - L\textbf{I}_{M}||^{2} + 2\varSigma_{n=1}^{L-1}||\textbf{R}_{n}||^{2},
\end{equation}
where $\textbf{R}_{n}$ is the coded waveform covariance matrix at lag n, $L$ is the length of the code and $\textbf{I}_{M}$ is an identity matrix of size $M$. $M$ signify the number of sequences in the code set. For our case, $M = 2$. This error function can be represented conveniently in the spectral domain, to form a computationally efficient algorithm called CAN. The correlation function obtained from two sequences generated with CAN algorithm ($L$ = 256 samples) is shown in Fig. \ref{fig_sim3_1}. \par

Another version of the CAN algorithm is the weighted CAN (WeCAN) \cite{Haohe2009}, which aims at minimizing the following error function:
\begin{equation}
\varepsilon = \gamma_{0}^{2}||\textbf{R}_{0} - L\textbf{I}_{M}||^{2} + 2\varSigma_{n=1}^{L-1}\gamma_{n}^{2}||\textbf{R}_{n}||^{2},
\end{equation}
where ${\gamma}_{n=0}^{L-1}$ are real-valued weights. These can put emphasis on the reduction of sidelobe power in certain regions of the waveform ambiguity function. \par
A very significant point to be noted from \cite{Haohe2009} is that the $\varepsilon$ in both error functions of CAN and WeCAN, cannot be made arbitrarily small, even without the unit-modulus constraint on the absolute values of the code. But we can select a $P \leq (L+M)/2M$, which is a region around the mid-point on delay axis, of the correlation function. In that region $P$, it is possible to make $\varepsilon$ very small. This is depicted in the simulation of CAN derived code-filter pairs (Fig. \ref{fig_sim3_1}). These codes might be useful for applications like SAR imaging, where the transmit pulse lengths are large and we might be only interested in certain lags around the center of the correlation function. However, this performance is not enough for a weather radar applications, as the scatterers have a wide spatial and temporal distribution. 

\subsection{Mismatched-Filter-Based Orthogonal Polyphase Code:}\label{section2}
The motivation behind the use of mismatched-filter-based method is to have relatively larger filter lengths compared to the sequence length transmitted so that the peak sidelobe energy can be spread into much larger coefficient space. This approach has more flexibility as we can optimize the sequences and filters separately in an optimization framework, to obtain better sequences. We would start by design of optimal code-filter pair with regard to auto-correlation sidelobe energy (and later add cross-correlation sidelobe energy):

\begin{equation}
\textbf{y}_{ij} = \textbf{h}_{j}\textbf{X}_{i}^{m},
\end{equation}
where $\textbf{h}_{j}$ is the mismatched filter designed for the $j^{th}$ sequence in the orthogonal code set and $\textbf{X}_{i}^{m}$ is the modified transmit convolution matrix for the $i^{th}$ sequence, obtained by deleting the columns of transmit convolution matrix that corresponds to the mainlobe after convolution operation. The error function corresponding to the energy in auto-correlation sidelobes can be written as:

\begin{equation}
\varepsilon_{hi} = \textbf{h}_{i}^{t}\textbf{X}_{i}^{m}\textbf{X}_{i}^{mH}\textbf{h}_{i}^{*}.\\
\end{equation}
With the constraint of $\textbf{x}_{i}^{t}\textbf{h}_{i} = L$, $\textbf{x}_{i}$ being the $i^{th}$ sequence of the orthogonal code set and $L$ is the length of code, the error function has got a minimum at \cite{Griep1995}:

\begin{equation}
\textbf{h}_{i}^{t} = (\textbf{X}_{i}^{m}\textbf{X}_{i}^{mH})^{-1}\textbf{x}_{i}L / \textbf{x}_{i}^{H}(\textbf{X}_{i}^{m}\textbf{X}_{i}^{mH})^{-1}\textbf{x}_{i}.
\end{equation}
If the polyphase code-filter pairs to be optimized for both auto- and cross-correlation energy (using mismatched filter), then the error function gets modified as:
\begin{equation} \label{eq_12}
\varepsilon_{hi} = \textbf{h}_{i}^{t}\varSigma_{i=1}^{M}(\textbf{X}_{i}^{m}\textbf{X}_{i}^{mH})\textbf{h}_{i}^{*},
\end{equation}
where $M$ is the number of codes in the orthogonal set and the optimum filter weights would be:
\begin{equation} \label{eq_13}
\textbf{h}_{i}^{t} = (\varSigma_{i=1}^{M}(\textbf{X}_{i}^{m}\textbf{X}_{i}^{mH})^{-1}\textbf{x}_{i}L / \textbf{x}_{i}^{H}(\varSigma_{i=1}^{M}(\textbf{X}_{i}^{m}\textbf{X}_{i}^{mH}))^{-1}\textbf{x}_{i}.
\end{equation}

This is the well-known closed form expression for the minimization vector:
\begin{equation} \label{eq_14}
\textbf{h}_{i}^{t} = (\varSigma_{i=1}^{M}(\textbf{X}_{i}^{m}\textbf{X}_{i}^{mH}))^{-1}\textbf{x}_{i}, 
\end{equation}
and the denominator part of equation \eqref{eq_13} is for normalization.

Our method starts with a set of two codes, randomly generated. For simulation, we took $L=40$ for a pulse width of $20\mu s$ and bandwidth of 2 MHz. The filters based on equation \eqref{eq_14}  were generated (optimal in ISL of the auto- and cross-correlation sidelobes). After we have obtained a set of initial codes and filters, then we iterate over the permissible code combinations and simultaneously trying to find minima for the error function in equation \eqref{eq_13} using optimization methods. In the next iteration, when a new code is found, again we get ISL filter using equation \eqref{eq_14} and the error function is evaluated. This continues until we reach at least a local minima. The mainlobe width is assumed to be 5 samples. This method jointly optimizes the polyphase codes and filters. Subsequently, global optimization forces the local solvers to run from various initial points in the search space, which are obtained using the scatter search. More details in \ref{section_optim} about setting up the optimization process. \par
It is easy to realize that it is difficult to obtain very low levels of auto-correlation and cross-correlation of sequences. In reality, there are bounds on how low the correlation sidelobes can be simultaneously achieved. Initially, it was explored by Dr. Sarwate in \cite{Sarwate1979} and Welch in \cite{Welch1974}. Moreover, the potential application of sequences with good correlation properties is immense, for example, in a multiple-input multiple-output (MIMO) radar \cite{Stoica2008}, communication, etc. When we employ orthogonal signals for transmitting in a MIMO radar, there is a significant improvement in detection performance due to spatial diversity. Also, these codes can significantly improve the bandwidth efficiency of a code-division multiple-access (CDMA) based system in addition to enhancing the interference-free operation \cite{Liu1995}. In the weather radar community, such codes can potentially lead to instantaneous measurement of scattering matrix parameters \cite{Wang2010}. Another class of polyphase codes that try to achieve orthogonality of sequences is complete complementary codes (CCC). They are generalizations of space-time block codes and Golay complementary codes, developed to explore polarization diversity. Golay pairs have zero auto-correlation sidelobes if the autocorrelation function of the pair is added together, which essentially means that the pair has complementary sidelobes. Additionally, CCC's have the property that its cross-correlation sidelobes also vanishes for the sum of cross-correlations of the pairs in the set. A very general definition of CCC is given in \cite{Tang2014}. Much finer details about their construction methods and making these codes doppler tolerant are given in \cite{Pezeshki2008} -  \cite{Han2011}. \par

Another point to be noted here is that, for example, in D3R radar, the number of mismatched filter coefficients is fixed (based on resources) but the pulse width might vary, for example, for a 1MHz bandwidth, we have $20 \mu s$ or $40 \mu s$ pulse width. However, since the ratio of the number of waveform coefficients to the number of filter coefficients (WF2FC) has become twice (in case of $40 \mu s$ pulse width), we have observed peak sidelobe levels degrading by as much as 10 dBs. The worst case peak sidelobe levels are obtained when the number of waveform coefficients becomes equal to the number of filter coefficients, which is a case of matched filter design. Additionally, if the bandwidth increases, even with the same pulse width, the number of waveform coefficients would increase and the ratio of WF2FC would increase as well, having an adverse impact on the peak sidelobe ratios. Thus this technique works best when the WF2FC ratio is small (0.042 for D3R operating at 1MHz bandwidth and $20 \mu s$ pulse width).

\subsection{Setting up the Optimization Problem} \label{section_optim}
The goal of the optimization problem is defined by the objective function, $g(x)$, and a set of constraints on the variables, which define the limits of acceptable space for these variables. The standard problem is defined by:

\begin{equation} \label{eq_14}
\begin{aligned}
&min_{x} g(x)\\
&s.t. E_{i}(x) = 0, i=1,...,m\\
&I_{j}(x) \leqslant 0, j=1,...,n\\
&\textbf{x}_{l} \leqslant \textbf{x} \leqslant \textbf{x}_{u},
\end{aligned}
\end{equation}
where $E_{i}$ and and $I_{i}$ are equality and inequality constraints for $\textbf{x}$. The last statement tells us about the lower and upper bounds on $\textbf{x}$. A global optimizer is the one that finds  $\textbf{x}_{min}$ from the full feasible set. In case, either the objective function or the constraints are non-linear then such a problem can be shown to have first-order optimal solution, if it satisfies the Karush Kuhn Tucker (KKT) conditions \cite{Tucker1970}, through the use of Lagrange's multipliers. These are necessary to combine the objective function and the constraints and they scale the gradients to fulfill the condition. The optimization problem of minimal sidelobe energy in equation \eqref{eq_13} is a non-linear function with non-convex unit modulus constraint, which defines the $g(x)$ and $E_{i}(x)$ of equation \eqref{eq_14}. The inequality constraints and the bounds on $x$ are defined in section \ref{sec_cons}. Hence the combined objective function satisfying the KKT conditions will be non-convex and non-linear having multiple local minima. For such a case, we need to run global optimization routines to find an optimal solution. As the pulse width or the bandwidth increases, we would have more variables to optimize with each variable able to take valid values between $[0,2\pi]$. It can be observed that the search space is huge and the global minima will be difficult to obtain with direct search methods. Hence the use of gradient-based method.\par

The gradient-based method requires that the objective function, constraints, and gradients be continuous. We have used Sequential Quadratic Programming (SQP) and Active set based gradient search. These methods transform the given problem to a set of sub-problems that may be easier to solve. These are likely to converge to a local minima for a non-convex non-linear constraint problem and require gradient and Hessian. The gradient, if not supplied, is approximated using finite difference method and the Hessian is usually computed at each iteration using a quasi-Newton method called Broydon-Fletcher-Goldfarb-Shanno (BFGS) \cite{Fletcher1970}. \par

Both multi-start or global search can be used along with a local solver over the whole of search space, for our non-convex non-linear problem and constraint to obtain a global minima. Multi-start runs the local optimizer from multiple feasible start points. Individual runs can be in parallel. It starts with uniformly distributed start points within the domain. Global search, however, starts with initial point $x_{0}$ and a set of potential starting points using scatter search \cite{marti2006}. Since global search analyzes start points and reject those points that are not likely to improve upon the best minima, it is faster and more logical to work with. In section \ref{section_derivative}, we derive the gradient of the error function so that local search can likely converge faster. \par

The computational search solvers have been studied extensively like the use of exhaustive search, evolutionary algorithms and heuristic search, for the design of polyphase codes or sequences. They can all be used to optimize the local search around an initial point. The exhaustive search complexity is $O(2^{n})$ for binary sequences of length n and it can be reduced to  $O(1.85^{n})$ by using branch and bound \cite{Mertens1996}. However, this complexity would increase further with the number of symbols in the polyphase sequence hence heuristics methods such as \cite{Tan2016} should be used for large sequences. The evolutionary methods such as genetic algorithms are likely to fail because the generation step creates candidates by mutation or crossover. But a linear combination of two good sequences will not necessarily provide a sequence that has better correlation properties. The cyclic algorithm named CAN, introduced in \cite{Haohe2009}, provides sequences of larger lengths but instead of the original ISL cost function, it tries to minimize another simpler criterion which is shown to be equivalent to ISL. However, it needs to be modified to take our framework of mismatched filter (CAN based algorithms are derived with the matched filter approach). Even the majorization-minimization (MM) method in \cite{Song2016}, which works directly with the ISL cost function has to be modified significantly for use with mismatched filter.\par

\subsection{Constraints} \label{sec_cons}
We already mentioned the uni-modular constraint on the optimization problem. This is important for the transmit section which might need to operate in saturation mode. Moreover with the use of a mismatched filter, an overall constraint on the gain and phase terms of the combined code-filter pair has to be incorporated. Apart from this, there is also a need for another constraint that the gain and phase terms of code-filter pair 1 be equal to gain and phase terms of code-filter pair 2. These pairs were obtained using the joint optimization procedure described in section \ref{section_optim}. This would ensure that for the velocity retrieval, the phase imbalance of odd and even pulses, do not give rise to additional spurious velocity components. The gain balance between these pairs would aid in reflectivity calibration.

\subsection{Derivative of the Error/Energy function} \label{section_derivative}
Let us look first, at the gradient of autocorrelation sidelobe energy (or the error function). If $\textbf{a}$ is the polyphase sequence with unit magnitude constraint, then
\begin{equation}
c_{m} = \varSigma_{i=1}^{N}a_{i}a^{*}_{i+m}, m \in (-N,N).
\end{equation}

With the weighing and exponential term in the error function, as used in \cite{Baden2015}:
\begin{equation}\label{eq_16}
\varepsilon = 2\varSigma_{m=1}^{N}w_{m}(c_{m}c^{*}_{m})^{p}.
\end{equation}
It turns out that the partial derivative of the error function w.r.t. $\alpha$, which is the phase angle of  $\textbf{a}$, comes out to be \cite{Baden2015}:
\begin{equation}
\delta = \frac{\partial \varepsilon}{\partial \alpha} = 4p \Im(\textbf{a} \circ ((\boldsymbol{\beta} \circ \textbf{c}) * \textbf{a})),
\end{equation}

where $\beta_{m} = w_{m}(c_{m}c^{*}_{m})^{p-1}$ and $x \circ y$ is the Hadamard product of x and y.\par

Next for our application, we expand this derivative for a mismatched filter case and we begin with:
\begin{equation}
c_{m} = \varSigma_{i=1}^{N}a_{i}b^{*}_{i+m}, m \in (-N,N),
\end{equation}
where $\textbf{b}$ contains the filter coefficients and N is its length. After forming the error function as per equation \eqref{eq_16} and taking its derivative w.r.t. real part of $\textbf{a}$, we get,
\begin{equation}
\begin{aligned}
\frac{\partial \varepsilon}{\partial \Re(a_{j})} &= 2p\varSigma_{m=-N+1}^{N-1}\beta_{m}[\Re(c_{m}) \frac{\partial \Re(c_{m})}{\partial \Re(a_{j})} + \Im(c_{m}) \frac{\partial \Im(c_{m})}{\partial \Re(a_{j})}]\\
\Re(c_{m}) &= \Re(\varSigma_{i=1}^{N} a_{i}b^{*}_{i+m})\\
&= \varSigma_{i=1}^{N} [\Re(a_{i})\Re(b_{i+m}) + \Im(a_{i})\Im(b_{i+m})]\\
\Im(c_{m}) &= \varSigma_{i=1}^{N} [-\Re(a_{i})\Im(b_{i+m}) + \Im(a_{i})\Re(b_{i+m})].\\
\end{aligned}
\end{equation}

The partial derivatives are zero unless $i=j$, so that,

\begin{equation}
\begin{aligned}
&\frac{\partial \Re(c_{m})}{\partial \Re(a_{j})} = \Re(b_{j+m})\\
&\frac{\partial \Im(c_{m})}{\partial \Re(a_{j})} =  -\Im(b_{j+m}).\\
\end{aligned}
\end{equation}

Thus, we can write,
\begin{equation}
\begin{aligned}
\frac{\partial \varepsilon}{\partial \Re(a_{j})} = 2p\Sigma_{m=-N+1}^{N-1}\beta_{m} [\Re(c_{m})\Re(b_{j+m}) - \Im(c_{m})\Im(b_{j+m})].
\end{aligned}
\end{equation}

Similarly, we can write,
\begin{equation}
\begin{aligned}
\frac{\partial \varepsilon}{\partial \Im(a_{j})} = 2p\Sigma_{m=-N+1}^{N-1}\beta_{m} [\Re(c_{m})\Im(b_{j+m}) - \Im(c_{m})\Re(b_{j+m})].
\end{aligned}
\end{equation}
Let $\alpha$ be the phase angle of $\textbf{a}$, then using the chain rule we get,
\begin{equation}
\frac{\partial \varepsilon}{\partial \alpha_{j}} = \Re(a_{j})\frac{\partial E}{\partial \Im(a_{j})} - \Im(a_{j})\frac{\partial E}{\partial \Re(a_{j})}.
\end{equation}
Using equations above, we get,
\begin{equation}
\begin{aligned}
\frac{\partial \varepsilon}{\partial \alpha_{j}} &= 2p\Sigma_{m=-N+1}^{N-1}\beta_{m} [\Re(c_{m})\Im(b_{j+m})\Re(a_{j}) \\
&- \Im(c_{m})\Re(b_{j+m})\Re(a_{j}) - \Re(c_{m})\Re(b_{j+m})\Im(a_{j}) \\
&+ \Im(c_{m})\Im(b_{j+m})\Im(a_{j})].
\end{aligned}
\end{equation}

This derivative of the error function, is a guiding direction in the local search under framework of codes and mismatched filters.

\subsection{Ambiguity Function for mismatched filters}
The ambiguity function for a mismatched filter represents the effect of delay and doppler on sidelobes and mainlobe, on a code-filter pair. In general, ambiguity functions are good tools to analyze the performance of synthesized polyphase codes for doppler tolerance in a weather sensing application. In any case, the ambiguity function at $(\tau,f_{d}) = (0,0)$ corresponds to a matched output to the signal reflected perfectly from the target of interest. The ambiguity function $|\chi(\tau,f_{d})|$ for a polyphase code $\textbf{a}$ and mismatched filter $\textbf{b}$, can be written as:
\begin{equation}
|\chi(\tau,f_{d})| = |\varSigma_{n=-N+1}^{N-1} a(n)b(n+\tau) exp(j2\pi f_{d}n)|.
\end{equation}

In plots, we would be showing only the positive Doppler shifts as the ambiguity function is symmetric around zero. We start with ambiguity functions achieved with optimization carried out for auto-correlation sidelobe energy, as the error function (using mismatched ISL filters). 

\begin{figure}[!t]
	\centering
	\includegraphics[width=3in]{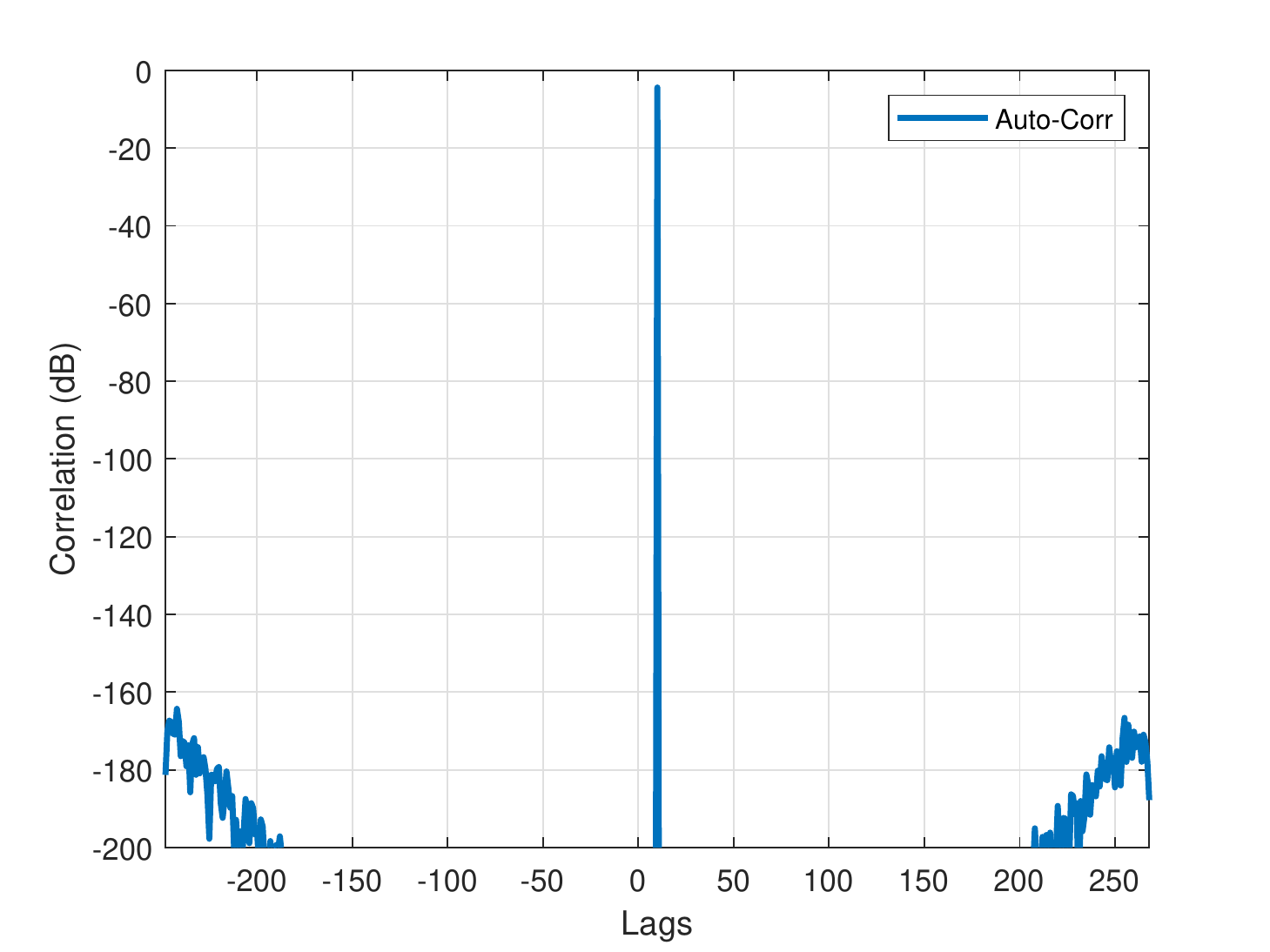}
	\caption{The zero doppler cut of the synthesized polyphase code of length 40 with uni-modular constraint (from auto-ambiguity function).}
	\label{fig_synthcode}
\end{figure}

\begin{figure}[!t]
	\centering
	\includegraphics[width=3in]{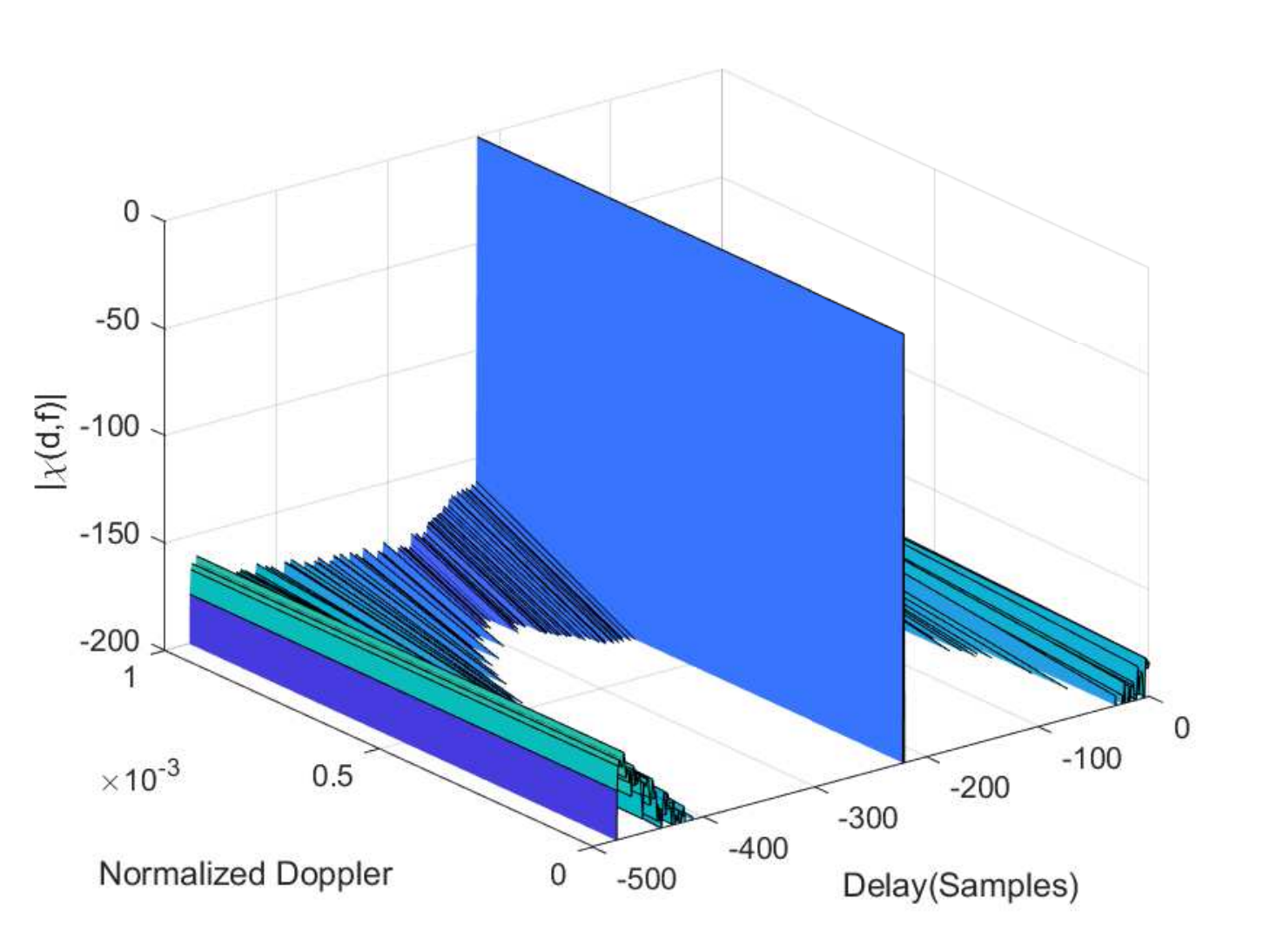}
	\caption{Auto-ambiguity Function of polyphase code with length 40 samples (PW = $20 \mu s$, BW = 2MHz) and the mismatched filter length of 480 samples.}
	\label{fig_ambcode}
\end{figure}
\begin{figure}[!t]
	\centering
	\includegraphics[width=3.5in]{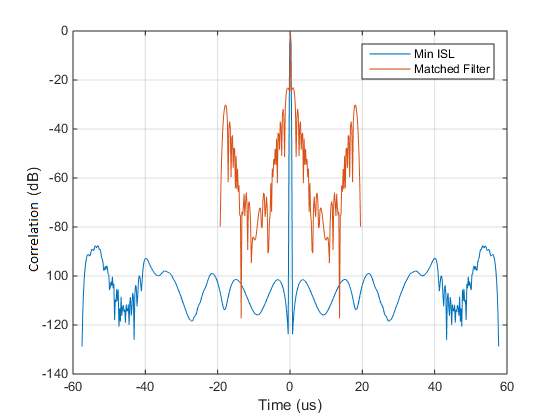}
	\caption{A comparison of peak sidelobe levels for a matched and mismatched-filter-based on Chirp waveform. Blue depicts mismatched filter designed with minimum ISL criterion}
	\label{fig_comp}
\end{figure}

Fig. \ref{fig_synthcode} and \ref{fig_ambcode} show the zero-doppler cut (from ambiguity function) of a polyphase code and filter with uni-modular constraint (synthesized with only auto-correlation sidelobe energy error function with global search). It used mismatched filter length of 480 coefficients. The zero Doppler cut has very low sidelobes in the whole domain with a one-sample mainlobe. The doppler performance of the code is also good under reasonable doppler assumption (peak sidelobe level below -150dB). The sidelobes are practically zero throughout the correlation space for zero doppler condition. Such level of sidelobe performance is far superior to the matched filter counterparts and even for Chirp based mismatched filters. Both matched filter and mismatched filter sidelobe level for a Chirp based waveform is depicted in Fig. \ref{fig_comp}. The peak sidelobe level for a Chirp based mismatched filter is $\sim -85dBc$ (the $dBc$ units used here is the relative power of the peak sidelobes with respect to the carrier power normalized to 0 dBm). However, the polyphase based optimization yields peak sidelobe performance better than $-180dBc$. \par

The Auto-ambiguity and the Cross-ambiguity functions, are defined next, in the context of our framework. Let the orthogonal code set comprise of two polyphase codes, $\{\textbf{a}_{1},\textbf{a}_{2} \} \in \mathbb{C}$ and the corresponding ISL filters be $\{\textbf{b}_{1},\textbf{b}_{2} \} \in \mathbb{C}$. These sets are obtained minimization of auto-correlation sidelobe energy in $\{\textbf{a}_{1} * \textbf{b}_{1} + \textbf{a}_{2} * \textbf{b}_{2} \}$ and cross-correlation sidelobe energy in $\{\textbf{a}_{1} * \textbf{b}_{2} + \textbf{a}_{2} * \textbf{b}_{1} \}$. Under these assumptions, the auto-ambiguity function is defined as:

\begin{equation}
\begin{aligned}
&|\chi_{1auto}(\tau,f_{d})| = |\varSigma_{n=-N+1}^{N-1} a_{1}(n)b_{1}(n+\tau) exp(j2\pi f_{d}n)|,\\
&|\chi_{2auto}(\tau,f_{d})| = |\varSigma_{n=-N+1}^{N-1} a_{2}(n)b_{2}(n+\tau) exp(j2\pi f_{d}n)|.
\end{aligned}
\end{equation}
whereas the cross-ambiguity function is defined as:
\begin{equation}
\begin{aligned}
&|\chi_{1cross}(\tau,f_{d})| = |\varSigma_{n=-N+1}^{N-1} a_{1}(n)b_{2}(n+\tau) exp(j2\pi f_{d}n)|,\\
&|\chi_{2cross}(\tau,f_{d})| = |\varSigma_{n=-N+1}^{N-1} a_{2}(n)b_{1}(n+\tau) exp(j2\pi f_{d}n)|.
\end{aligned}
\end{equation}

\begin{figure}[!t]
	\centering
	\includegraphics[width=3in]{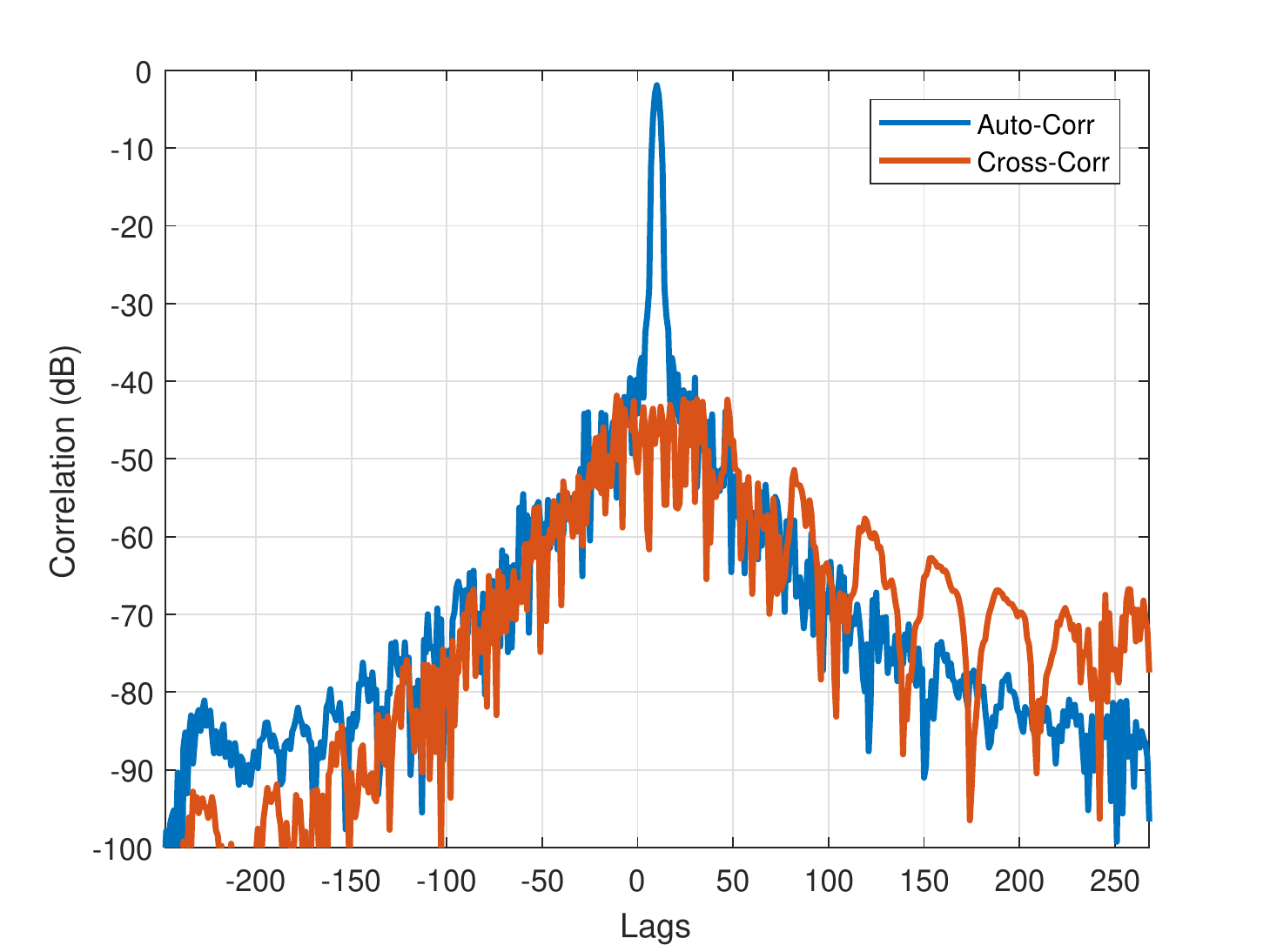}
	\caption{The zero doppler cut of the ambiguity function for the auto-correlation $\{\textbf{a}_{1} * \textbf{b}_{1}\}$ and cross-correlation $\{\textbf{a}_{1} * \textbf{b}_{2}\}$. The mainlobe width is set to 5.}
	\label{fig_zerodopp}
\end{figure}

\begin{figure}[!t]
	\centering
	\includegraphics[width=3in]{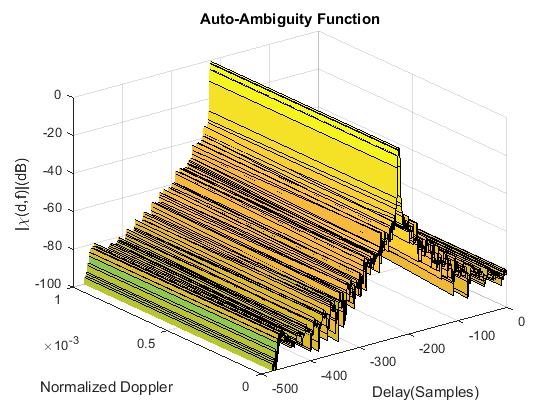}
	\caption{The Auto-ambiguity function plot of $\{\textbf{a}_{1} * \textbf{b}_{1}\}$.}
	\label{fig_autoamb}
\end{figure}

\begin{figure}[!t]
	\centering
	\includegraphics[width=3in]{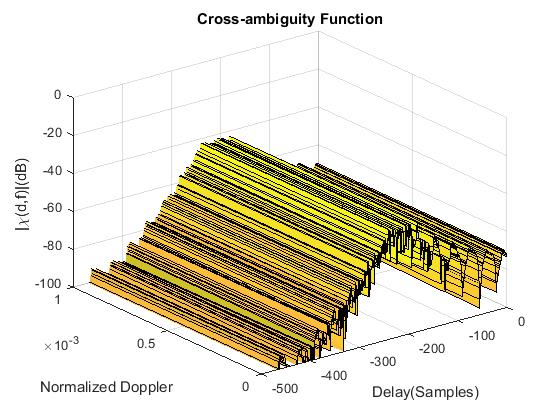}
	\caption{The Cross-ambiguity function plot of $\{\textbf{a}_{1} * \textbf{b}_{2}\}$.}
	\label{fig_crossamb}
\end{figure}

Let us see the physical realization of these with the help of ambiguity function plots for a pair of orthogonal polyphase codes and mismatched filters synthesized with global search. The advantages of global search were discussed in section \ref{section_optim} along with scatter search algorithm (to obtain good initial points for the local search). This is used here to come up with code-filter pairs. The error/cost function used is equation \eqref{eq_12} along with the constraints in section \ref{sec_cons}. Additionally, the derivative of error function in section \ref{section_derivative} is used to aid the optimization process. Finally, two code-filter pairs were obtained and their auto- and cross-correlation functions are depicted in Fig. \ref{fig_zerodopp} (zero-doppler cut). Fig. \ref{fig_autoamb} and \ref{fig_crossamb} gives the effect of moderate doppler frequencies on these code-filter pairs.

It can be easily observed from these plots that both the peak auto- and cross-correlation sidelobe levels are better than $-40dBc$, and in general, they are better than $-70dBc$ for half of the correlation space. This level of performance has not been demonstrated before, even for large code lengths and it can suppress the second trip echoes in a weather radar system to a great degree. Even though this is not sufficient for second trip retrieval but these codes work well for the suppression scheme as also substantiated from D3R weather radar data. Before implementing it on the actual system, we test it using a weather radar data simulator.

\section{Simulation of weather radar signals and modeling of phase noise errors} \label{section_3}
The computer simulation of In-phase and Quadrature returns from a weather scenario is essential to study the effectiveness of these coding techniques. With the simulator, we have the capacity to evaluate the newly developed polyphase code for weather radar in a controlled environment without actually physically testing on the radar platform. The parameters for simulation are the values from the NASA D3R weather radar system (Ku band) which has recently been upgraded with new IF sub-systems \cite{kumar2017}, \cite{kumar2018}. A brief specifications of this radar are given in Table \ref{table:specs}. The time series simulation of weather echoes for a radar is detailed in \cite{Choudhury2001}.

\begin{table}[ht]
	\caption{Brief Specifications of D3R (Ku band)} 
	\centering 
	\begin{tabular}{c c} 
		\hline\hline 
		Specification & Value \\ [0.5ex] 
		\hline 
		Frequency & 13.91 GHz $\pm$ 25 MHz  \\ 
		Range resolution & 150 m (nominal) \\
		Maximum operational range & 39.75 km\\
		Peak power & 200 W per H and V channel\\
		Receiver dynamic range &  90 dB \\[1ex] 
		\hline 
	\end{tabular}
	\label{table:specs} 
\end{table}

The weather radar simulator generates a Gaussian time series $x(nt_{s})$ with mean zero and variance $\sigma^{2}_{f}$. To impart a velocity $f_{D}$, 
\begin{equation}
y(nt_{s}) = x(nt_{s}) exp(2\pi f_{D}nt_{s}).
\end{equation}

\begin{figure}[!t]
	\centering
	\includegraphics[width=3.5in]{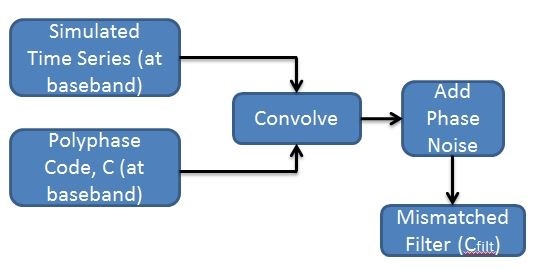}
	\caption{The process of simulating weather echoes to validate the performance of polyphase codes.}
	\label{fig_Simtimeseries}
\end{figure}

This time series is complex Gaussian centered at $f_{D}$. This received signal after down-conversion has a component of transmit and receive phase noise and can be written down as:
\begin{equation}
y(nt_{s}) = y_{wpn}(nt_{s})exp(j\phi(nt_{s})) + w(nt_{s}),
\end{equation}

where $\phi(nt_{s})$ denotes the phase noise process and $w(nt_{s})$ is zero-mean complex-valued additive white Gaussian noise (AWGN) that models the effect of noise from other components of the system. The phase noise can lead to perturbation of polyphase codes and the sidelobes. Convolution of simulated weather echo returns with polyphase code will lead to uncompressed receiver down-converter samples. Next, if we convolve it with mismatched filter coefficients, this would simulate polyphase code being transmitted, scattering through a weather event and being received/compressed with mismatched filter. Hence the utility of generated code-filter pairs can be assessed. It can be written down as:
\begin{equation}
y_{exp} = (y \ast C) \ast C^{*}_{filt},
\end{equation}
where $y_{exp}$ consists of samples obtained after effects of polyphase code $C$, mismatched filter coefficients, $C_{filt}$ and phase noise. The $\ast$ denotes convolution operator. The whole simulation process is depicted in Fig. \ref{fig_Simtimeseries}.

 \begin{figure}[!t]
	\centering
	\includegraphics[width=3.5in]{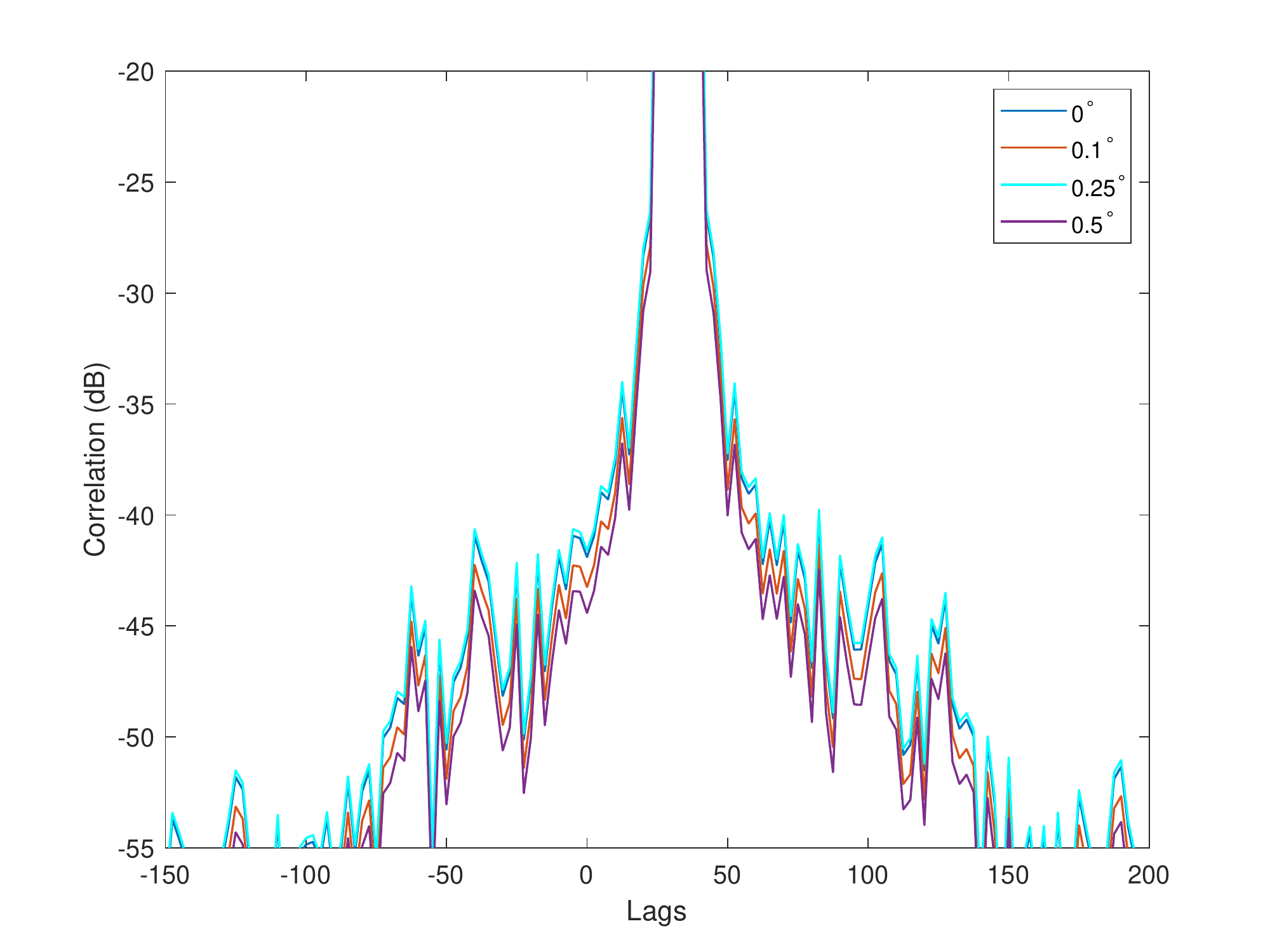}
	\caption{Effect of system phase noise on performance of code-filter pairs  with cumulative transmit and receive phase noise. The rms values shown are phase jitter calculated from phase noise, integrated over the receive bandwidth. }
	\label{fig_phasenoise1}
\end{figure}
\begin{figure}[!t]
	\centering
	\includegraphics[width=3.5in]{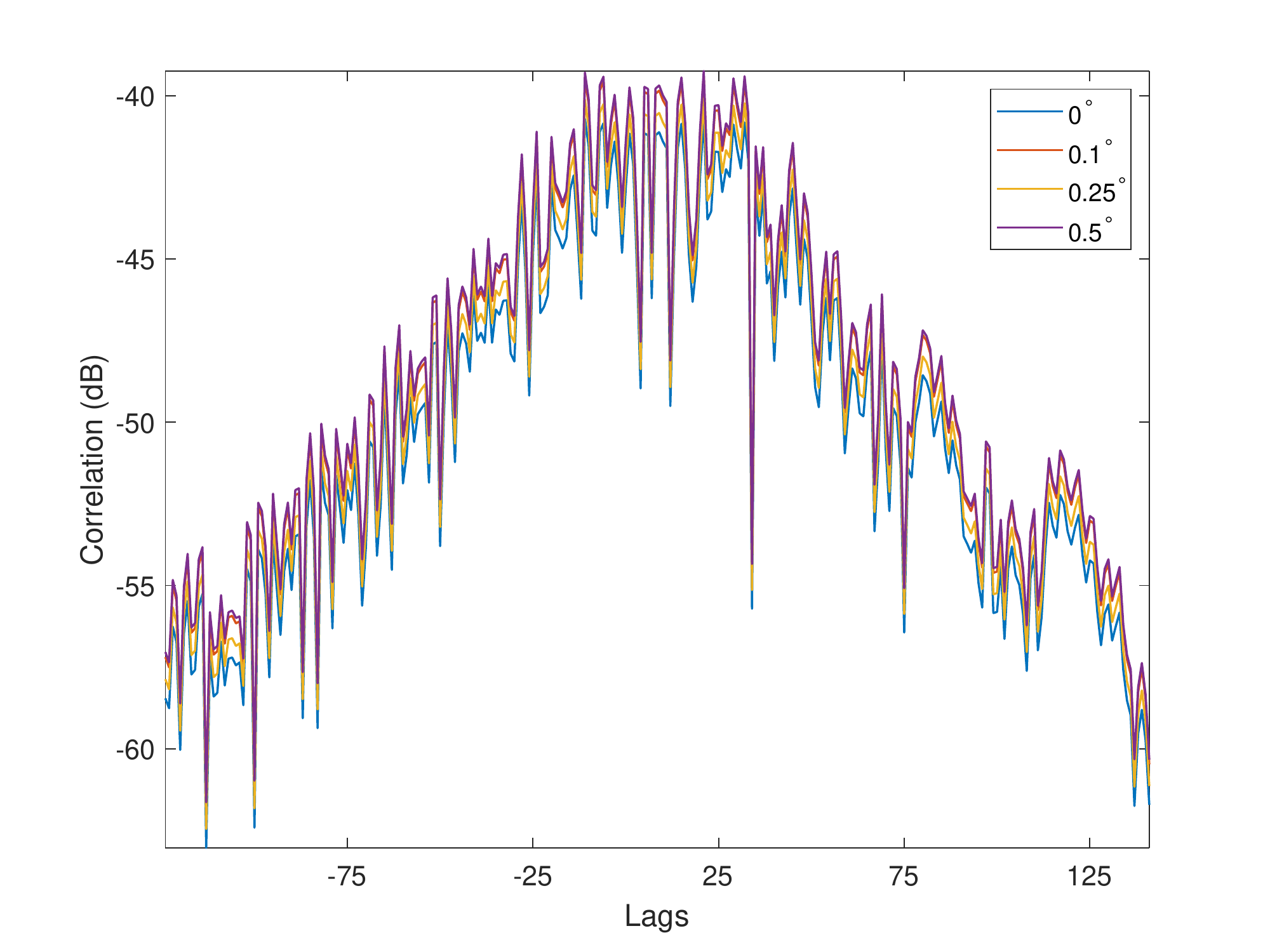}
	\caption{The phase noise degrades the peak sidelobe level in the cross-correlation function between the code 2 and mismatched filter 1. Only one pair of cross-correlation function is plotted here, to show the phase jitter effect.}
	\label{fig_phasenoise2}
\end{figure}

Before moving on to the actual weather time series simulation and results, we undertook a single target simulation to analyze the peak sidelobe degradation with phase noise. These are depicted in the Fig. \ref{fig_phasenoise1} and \ref{fig_phasenoise2}. The peak sidelobe level is getting worse by 3-4 dB as we go from no phase noise case to $0.5\degree$ RMS phase noise (phase jitter). Now let's proceed to define a method where we can incorporate the generated optimal two code-filter pairs and test for the second trip suppression capability of these codes.

\begin{figure}[!t]
	\centering
	\includegraphics[width=3.5in]{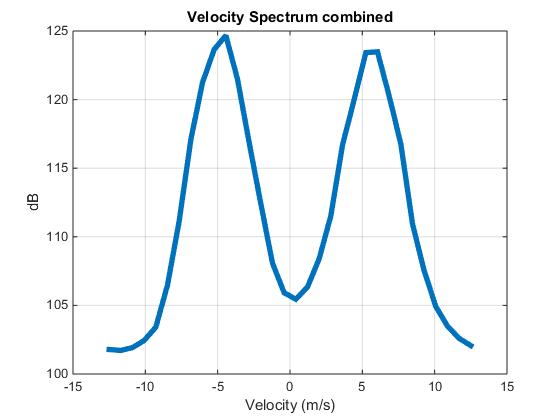}
	\caption{Overall combined spectrum of simulated first and second trip echo with equal power ratio with parameters: $BW=2MHz$, $PW=40\mu s$.}
	\label{fig_initCond}
\end{figure}
\begin{figure}[!t]
	\centering
	\includegraphics[width=3.5in]{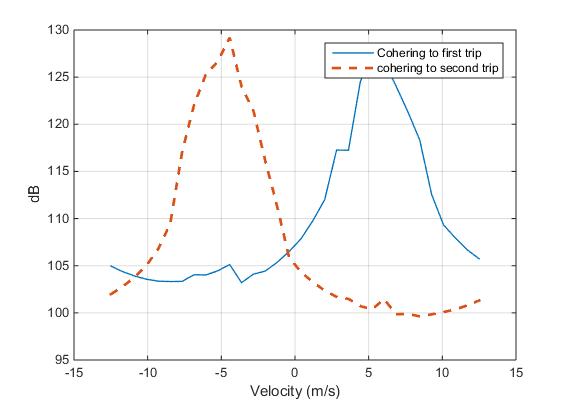}
	\caption{Second trip/first trip suppression obtained with orthogonal polyphase codes.}
	\label{fig_finalCond}
\end{figure}

\begin{figure*}[!t]
	\centering
	\begin{tabular}{c c c}
		\subfloat[]{\includegraphics[width=2in]{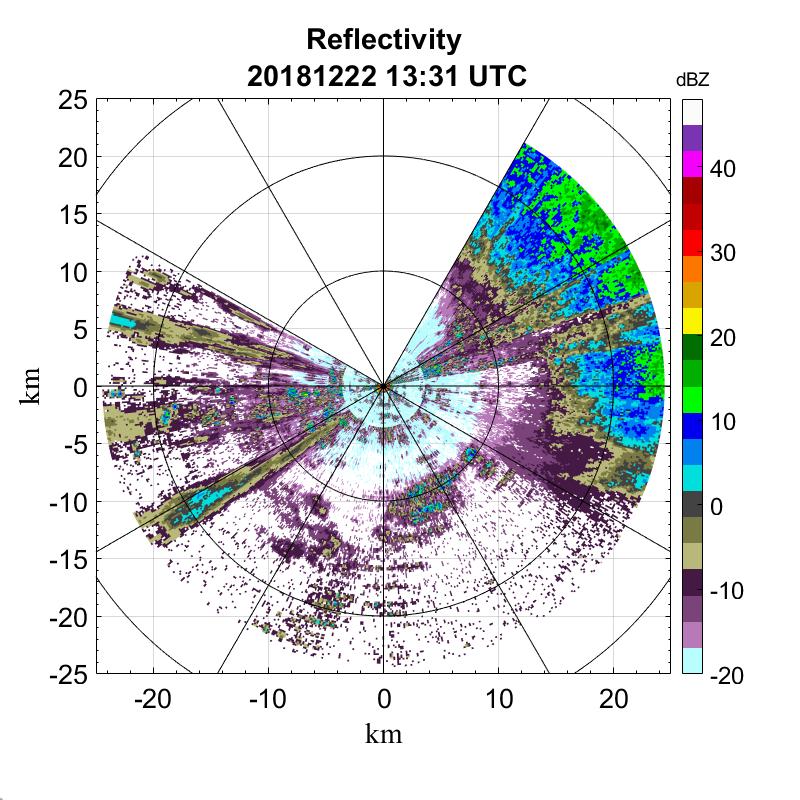}%
			\label{fig_first_case}}
		&
		\subfloat[]{\includegraphics[width=2in]{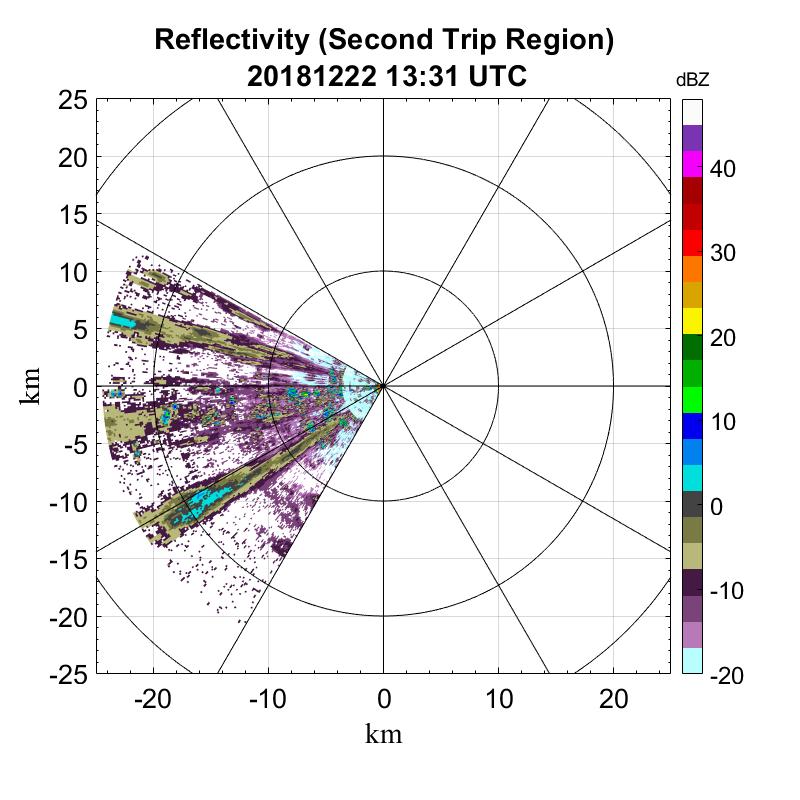}%
			\label{fig_second_case}}
		&
		\subfloat[]{\includegraphics[width=2in]{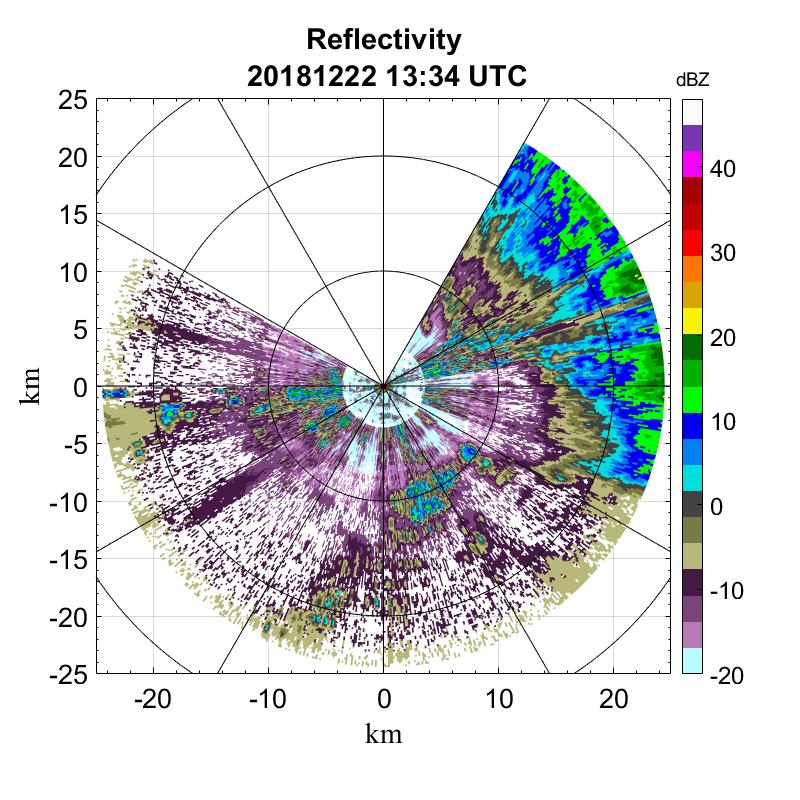}%
			\label{fig_third_case}}
		\\
		
		\subfloat[]{\includegraphics[width=2in]{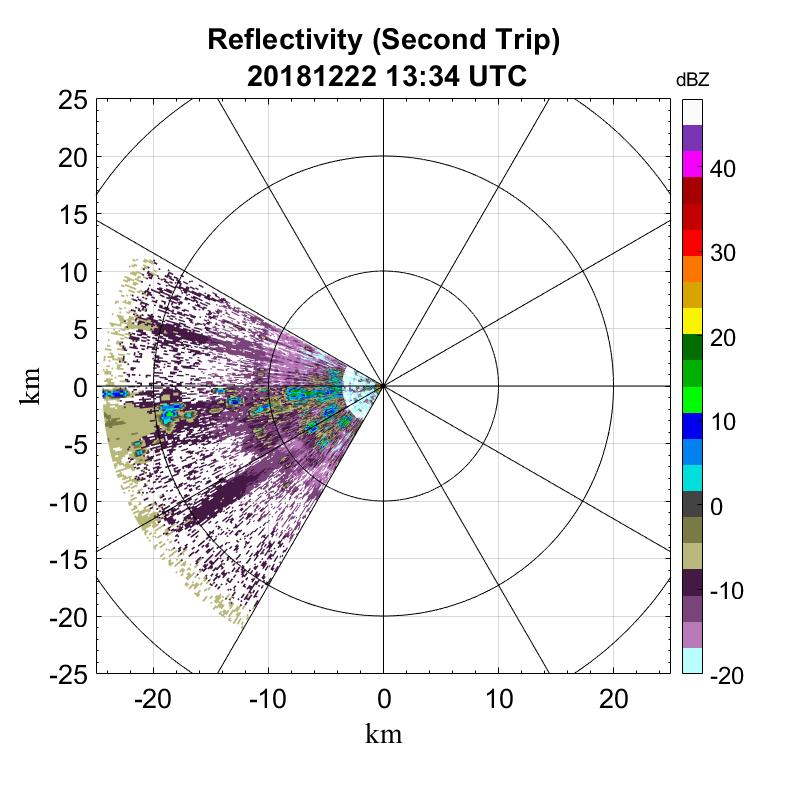}%
			\label{fig_fourth_case}}
		&
		\subfloat[]{\includegraphics[width=2.5in]{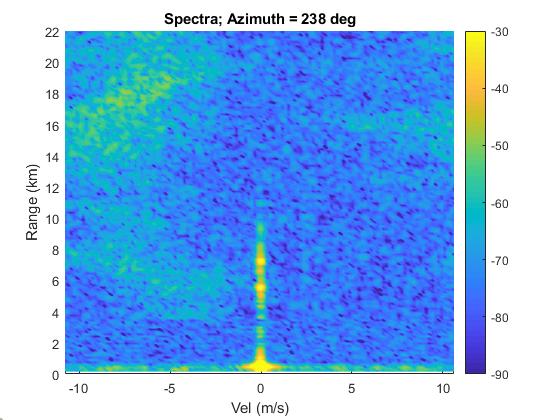}%
			\label{fig_fifth_case}}
		&
		\subfloat[]{\includegraphics[width=2.5in]{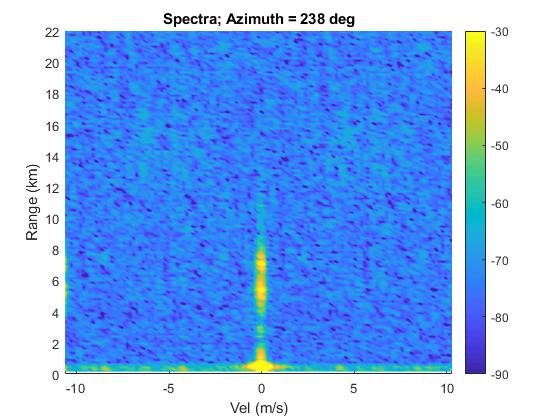}%
			\label{fig_sixth_case}}
		
	\end{tabular}
	\caption{(a) and (b) depict the reflectivity without polyphase codes. It is using a chirp waveform. The south-west region has second trips as confirmed with a Nexrad radar. The first trip lies towards the north-eastern region in this case. (c) and (d) are coded with orthogonal polyphase. The elevation is 2 deg and clearly suppression can be observed at $\sim$ radial 238 $\degree$ azimuth. (e) shows the doppler spectra along the same radial, for normal transmission, whereas, (f) has the same measurement but polyphase code and filter pairs are used.}
	\label{fig_case1}
\end{figure*}

\begin{figure*}[!t]
	\centering
	\begin{tabular}{c c}
		\subfloat[]{\includegraphics[width=2.5in]{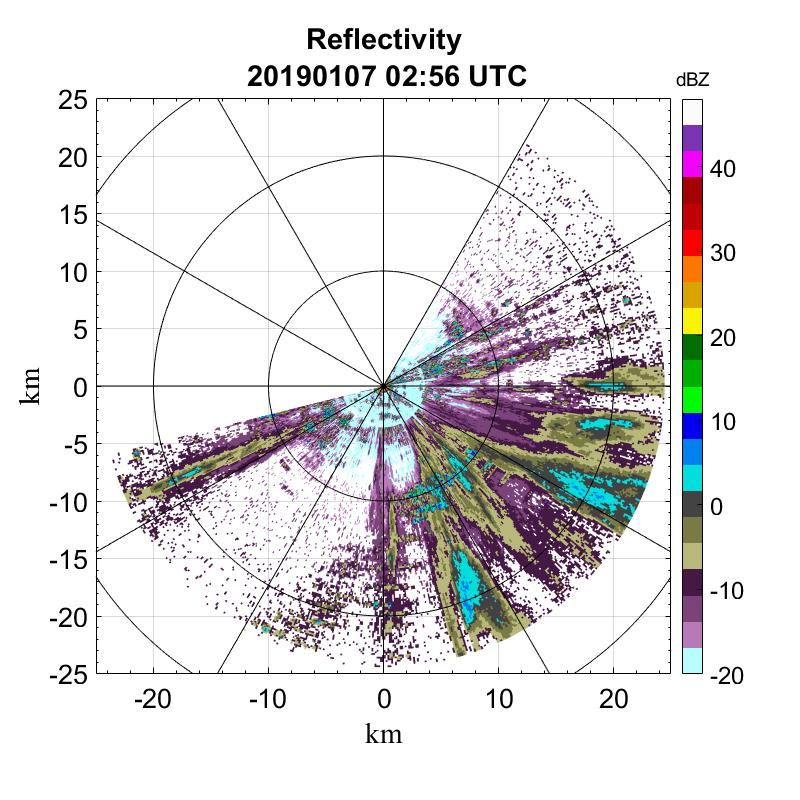}%
			\label{fig_first_case1}}
		&
		\subfloat[]{\includegraphics[width=2.5in]{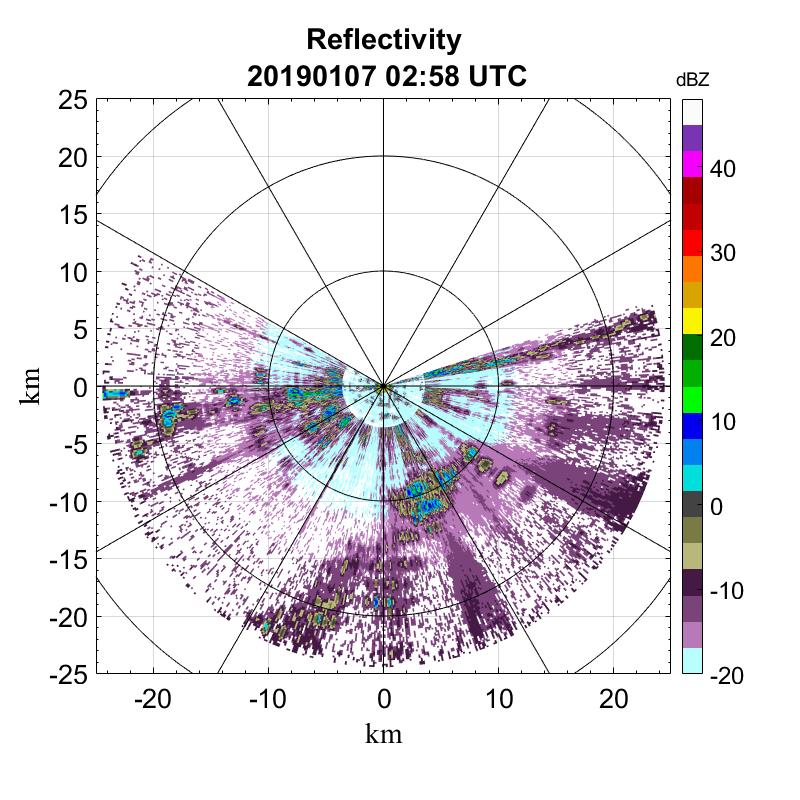}%
			\label{fig_second_case1}}
		
		\\
		\subfloat[]{\includegraphics[width=2.5in]{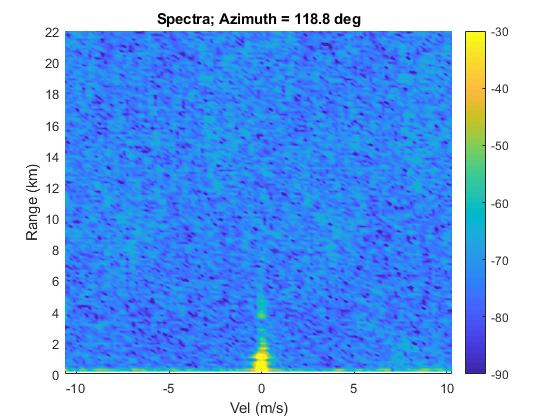}%
			\label{fig_third_case1}}
		&
		\subfloat[]{\includegraphics[width=2.5in]{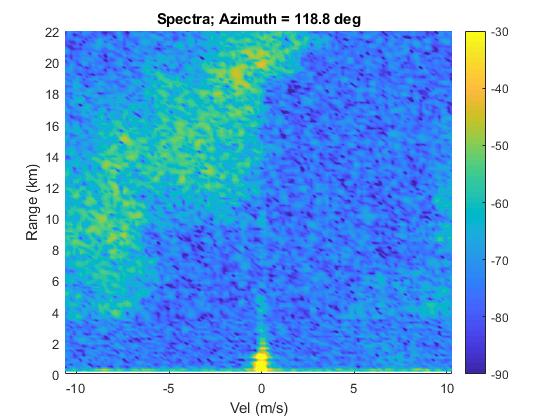}%
			\label{fig_fourth_case1}}
		
	\end{tabular}
	\caption{This case was recorded with D3R when only second trips are present in the unambiguous range.  (a) and (b) shows the reflectivity with normal transmission and with polyphase codes respectively. Reduction in second trip power are easily evident from these images. (c) and (d) are doppler spectra along a radial, for polyphase coded and normal transmission respectively.}
	\label{fig_case2}
\end{figure*}

\subsubsection{Method}
To test second trip suppression, one pair of code will be generated with good auto and cross-correlation
properties, by following the global optimization procedure in sections above. The alternate pulses will be modulated with the same code. Let's say, the pulse 1 gets modulated with code C1 and
the pulse 2 with code C2 and so on. The filters for this code to be loaded
in such a way that to recover first trip parameters, we load the filter with coefficients P1 and P2 during the pulse time-coded with C1 and C2 respectively. However, to recover the second
trip parameters, we load the coefficients P2 and P1 during the pulse time-coded with C1 and C2, respectively. It is to be noted that P1 was generated to optimize the autocorrelation with C1 and cross-correlation with C2 and vice-versa. To recover both trips simultaneously we need two filters, one loaded with coefficient sequence P1 and P2 for recovery of first trip while the other filter loaded with coefficient sequence P2 and P1 for recovery of second trip parameters. It is to be noted that in this paper, we are trying to show the suppression capability for second trip echoes and not their retrieval ability. A good retrieval code must have a much lower peak cross-correlation and sidelobe levels with the other pair (so that a much higher first trip power will have sufficient suppression, to be a practically viable option).

\subsubsection{Simulation Results} \label{sec_simRes}
To demonstrate improvements for second trip suppression, we convolve with the generated codes, pulse wise, to have the modulated time series echo returns from each range gate. Fig. \ref{fig_initCond} depicts the power ratios of the first and second trips used for simulation with the first trip centered at 5m/s while the second trip at -5m/s. After the combined signal goes through P1/P2 filters, Fig. \ref{fig_finalCond} depicts the power ratios achieved. There is $>20$dB of suppression of second trip observed using the simulation method and it can greatly reduce the bias on the measurement of dual-polarization moments of first trip. Hence the quality of measurements for the first trip can be vastly improved. Later similar levels of suppression of second trips were observed even in radar data.

\subsection{Effect of system phase on polyphase code performance}
\begin{figure}[!t]
 	\centering
 	\includegraphics[width=3.5in]{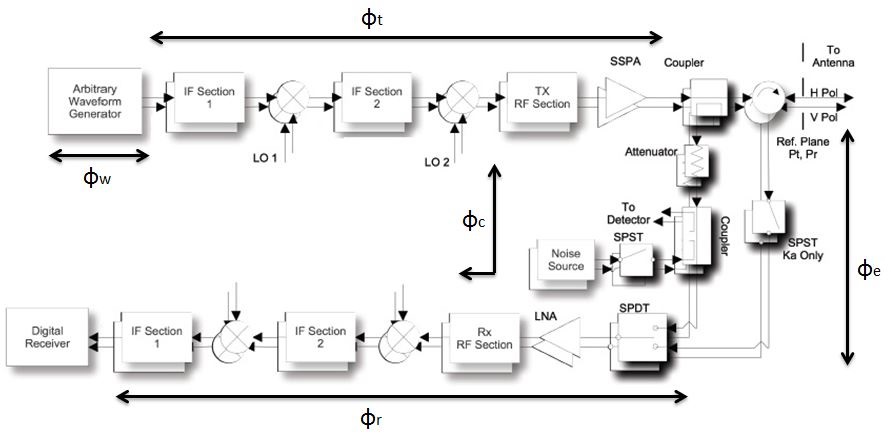}
 	\caption{The various constant system phase terms due to transmit components ($\phi_{t}$), receive components ($\phi_{r}$), the coupling elements in the calibration loop ($\phi_{c}$) and the extra path (different from calibration loop path) in the receive ($\phi_{e}$). For more details on D3R architecture, refer to \cite{Vega2014}.}
 	\label{fig_systemPhase}
 \end{figure}

There is an extra phase term added on top of the phase of the polyphase code which is due to the summation of group delay for the transmit and receive chain's components. This is referred to as the system phase and should be measured and subtracted out to improve performance. The true phase of the pulse compressor system using a polyphase generator can be obtained through simulation and the system phase can be measured on top of the true phase value, using the calibration loop. The various constant system phase terms in D3R architecture are shown in Fig. \ref{fig_systemPhase}. The $\phi_{w}$ is the actual code phase that should be ideally measured at the digital receiver. However, what would be measured is $\phi_{t} + \phi_{e} + \phi_{r} + \phi_{w}$, which are the transmitter phase, extra phase term and the receiver phase respectively. Through the calibration loop, we can measure $\phi_{t} + \phi_{c} + \phi_{r}$ where $\phi_{c}$ is the phase term through the coupler. This is the closest we can measure system phase in D3R and can be compensated.

\section{Observations from NASA D3R Weather radar} \label{section_4}
D3R has synchronous operation between Ku-band (13.91 GHz) and Ka-band (35.56 GHz). It is based on solid-state transmitters and the processing is capable of pulse compression in real-time using FPGA based digital
receiver system. Recently, the D3R radar was upgraded with a new version of digital receiver hardware and firmware which supports larger filter length and multiple phase coded waveforms and also newer IF sub-systems. The second trip for D3R would correspond to echoes beyond 75 km for a $500\mu s$ pri. One of the cases where suppression of second trips with the synthesized polyphase codes could be observed is depicted in Fig. \ref{fig_case1}. Although, we can have cross-correlation function (zero-doppler cut) $\sim$ 40dB below the auto-correlation peak, for a single target case. However, it should be emphasized that for a weather scenario, since the scatterers are continuous, the cross-correlation sidelobes may interfere constructively to degrade the peak sidelobe level achieved with a single target case.\par

\begin{figure}[!t]
	\centering
	\includegraphics[width=3.5in]{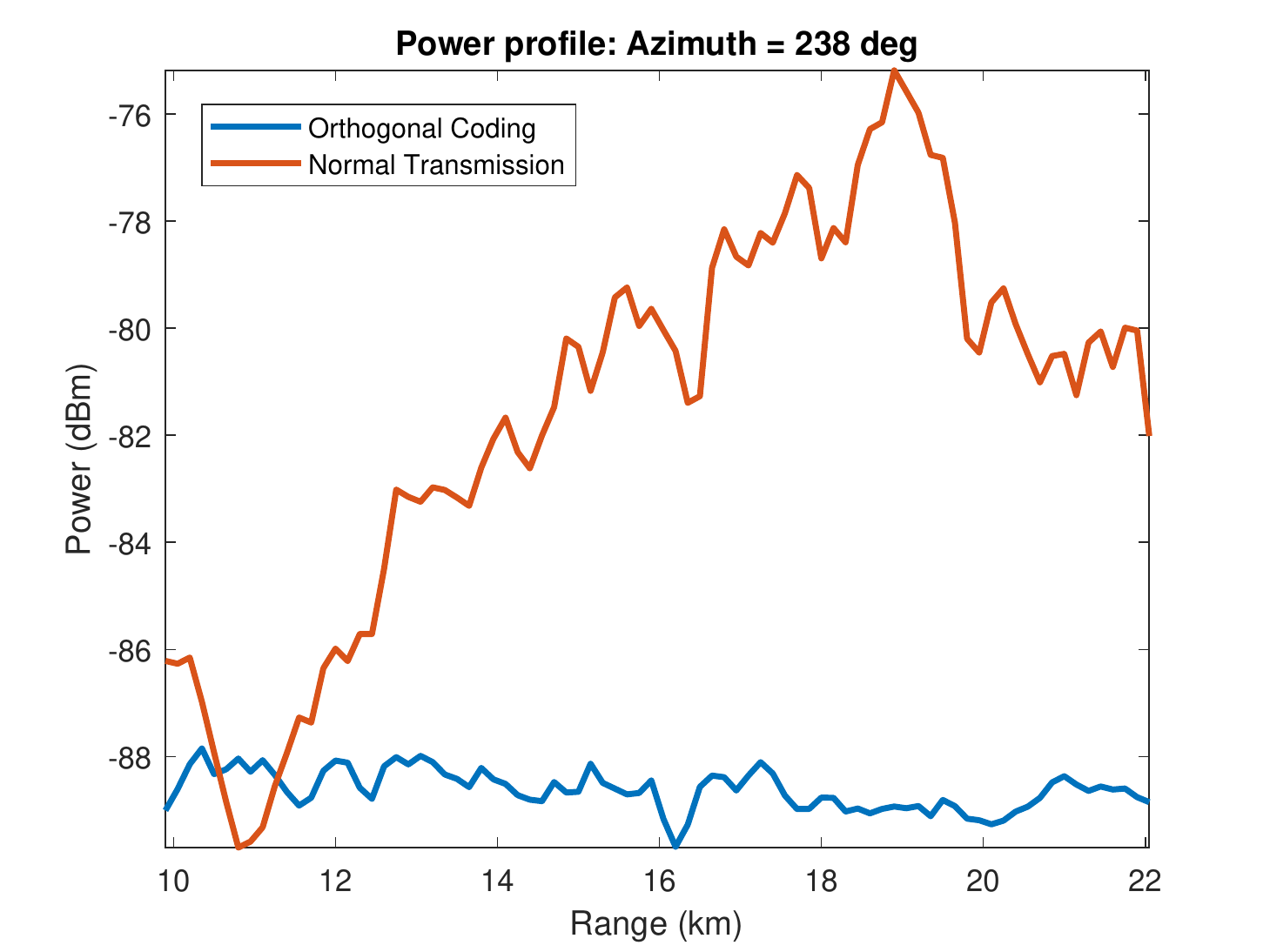}
	\caption{Power profile along ray at radial $238 \degree$ azimuth, clearly depicts the second trip suppression capability of the orthogonal polyphase codes.}
	\label{fig_powerProfile}
\end{figure}

\begin{figure}[!t]
	\centering
	\includegraphics[width=3.5in]{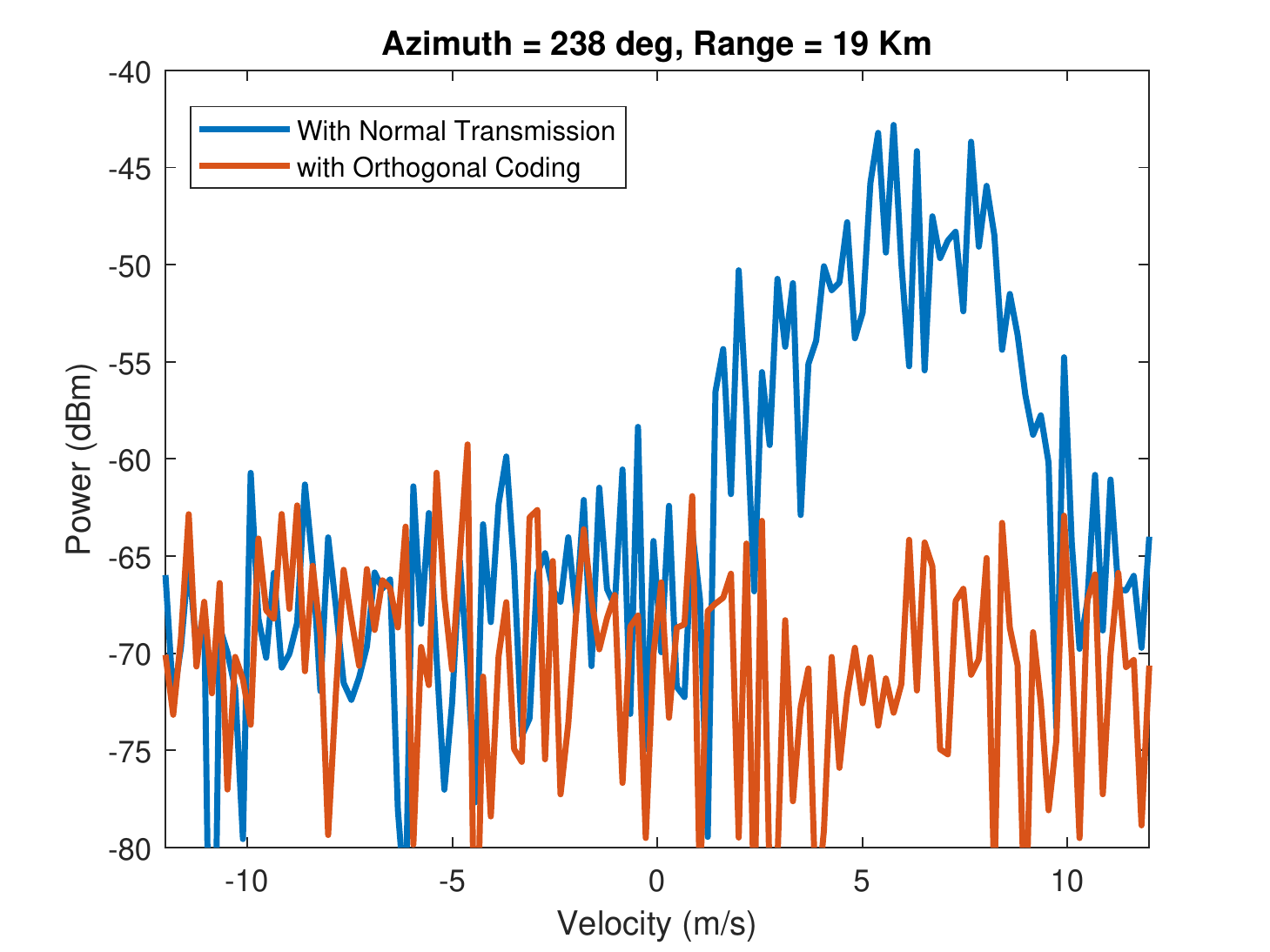}
	\caption{Velocity profile of second trip, observed for normal and polyphase coded transmission. These profiles are for cases shown in (b) and (d) of figure \ref{fig_case1}.}
	\label{fig_velPolt}
\end{figure}

A good amount of rejection of second trip echoes can be observed in this case, in the reflectivity and velocity spectrum, for the polyphase combination. To quantify this, we plot the power profile for the normal transmission and with polyphase coding for this case, along radial at $238 \degree$ azimuth. This is shown in Fig. \ref{fig_powerProfile}. It can be easily seen that the coding has cleared up the second trip echo and beyond 18 km of range, it is reduced below the noise floor. The velocity plot in Fig. \ref{fig_velPolt} compares the suppression that the polyphase codes have over normal transmission in the second trip velocity domain. More than 20dB of reduction in second trip power can be observed from these velocity plots. Another case with a second trip visible in the D3R unambiguous range is shown in Fig. \ref{fig_case2}. In this case, all the weather echoes appearing in radar range is second trip. Also, it was confirmed with nearby FTG radar station based in Denver, Colorado (part of NEXRAD weather network). The spectral plots for ray at radial $118.8 \degree$ azimuth, depicts the removal of second trip velocity. From these plots, at least 20dB of suppression of second trip echoes can be observed. And these observations are in line with the simulation results of section \ref{sec_simRes}. \par

The concept of intra-pulse coding (the polyphase coded waveforms demonstrated here) is quite different from the existing Chu code based suppression schemes (inter-pulse coding), which is being used in NEXRAD (Next generation weather radar) SZ modes. Some of these differences were enumerated in section \ref{Intro}. It should be emphasized that the second trip suppression capabilities of Chu inter-pulse based codes have been shown to be $\sim 40 dB$ in \cite{Zrnic1999} and in our case also, it is close to this number at similar phase noise conditions. However, the rejection of echoes from the trips of non-interest, in case of the Chu codes, is a function of spectral widths and would degrade in case
of multi-modal distributions and wider spectral width. 

\section{Conclusion}
The utility of intra-pulse polyphase coding techniques for weather radar systems is demonstrated and the recent research in developing codes with good correlation properties was utilized for weather radar application. A new polyphase code and mismatched-filter-based compression system has been developed to generate multiple code-filter pairs with optimal correlation properties. In this work, we focused on the design of two such pairs, however, in general, this framework can be used to synthesize multiple pairs. This technique has been applied for second trip suppression for D3R weather radar. Simulations were carried out, initially, to ascertain the performance of these new polyphase codes. The real-time implementation was carried out in NASA D3R weather radar, to see the suppression abilities of the developed code-filter pairs. Although the suppression ability of these codes are limited by the cross-correlation function between the pair of codes, shown to be $\sim$ 40 dB for single targets, a much better result could be obtained at the cost of mainlobe broadening (leading to reduced resolution) or a much larger length of the mismatched filter (more resources). We agree that a much higher level of performance is required for better suppression for second trip weather echoes and is a topic of further research. Additionally, the pair of codes used in this work, does not have suppression for odd trip echoes. However, using the same framework, more code-filter pairs can be synthesized, and third, fourth and so on pulses can be coded to achieve odd trip suppression as well. But the cross-correlation sidelobe levels may degrade.\par
Also, it should be highlighted that for a short-wavelength radar as D3R, a short duration ($1\mu$s) pulse is utilized to mitigate the blind range of the long duration ($20\mu$s) pulse. The short duration pulse is uncoded and would still have second-trip contamination. As far as dual-polarization moments like differential reflectivity, co-polar correlation coefficient, and differential phase are concerned, our idea is that since the same polyphase code-filter pair would be programmed on both horizontal and vertical polarizations, pulse wise. Then there is no adverse impact on these moments.\par

Another bottleneck lies in the implementation of large length of mismatched filters in firmware (FPGA) which complicates the overall design. But with the advancement of FPGA design techniques and better and larger devices available, this limitation is readily being overcome.


\bibliographystyle{IEEEtran}

\end{document}